\documentclass[a4paper,10pt,twocolumn, superscriptaddress, nofootinbib, amsmath,amssymb, aps, prd, floatfix]{revtex4-2}

\usepackage{graphicx}
\usepackage{amsfonts}
\usepackage{amsmath}
\usepackage{hyperref}
\usepackage{float}
\usepackage{amssymb,amsbsy}
\usepackage{epstopdf}
\usepackage{color}
\usepackage[normalem]{ulem}
\usepackage{braket}
\usepackage{combelow}

\def\bk{{\mathbf{k}}}
\def\feq{f_{0\mathbf{k}}}

\newcommand{\Kn}{\mathrm{Kn}}
\renewcommand{\Re}{\mathrm{Re}^{-1}}

\begin{document}

\title{
Transport coefficients of second-order relativistic fluid dynamics 
\\ in the relaxation-time approximation
}

\author{Victor E. Ambru\cb{s}}
\affiliation{Institut f\"ur Theoretische Physik, 
	Johann Wolfgang Goethe--Universit\"at,
	Max-von-Laue-Str.\ 1, D--60438 Frankfurt am Main, Germany}
\affiliation{Department of Physics, West University of Timi\cb{s}oara, \\
Bd.~Vasile P\^arvan 4, Timi\cb{s}oara 300223, Romania}

\author{Etele Moln\'ar}
\affiliation{Institut f\"ur Theoretische Physik, 
	Johann Wolfgang Goethe--Universit\"at,
	Max-von-Laue-Str.\ 1, D--60438 Frankfurt am Main, Germany}
\affiliation{Department of Physics, West University of Timi\cb{s}oara, \\
Bd.~Vasile P\^arvan 4, Timi\cb{s}oara 300223, Romania}
\affiliation{Incubator of Scientiﬁc Excellence--Centre for Simulations of Superdense Fluids,\\
University of Wroc{\l}aw, pl. M. Borna 9, PL-50204 Wroc{\l}aw, Poland
}

\author{Dirk H. Rischke}
\affiliation{Institut f\"ur Theoretische Physik, 
	Johann Wolfgang Goethe--Universit\"at,
	Max-von-Laue-Str.\ 1, D--60438 Frankfurt am Main, Germany}
\affiliation{Helmholtz Research Academy Hesse for FAIR, Campus Riedberg,\\
Max-von-Laue-Str.~12, D-60438 Frankfurt am Main, Germany}

\begin{abstract}
We derive the transport coefficients of second-order fluid dynamics with $14$ dynamical moments using the method of 
moments and the Chapman-Enskog method in the relaxation-time approximation for the collision integral of the 
relativistic Boltzmann equation. Contrary to results previously reported in the literature, we find that the second-order 
transport coefficients derived using the two methods are in perfect agreement.
Furthermore, we show that, unlike in the case of binary hard-sphere interactions,
the diffusion-shear coupling coefficients $\ell_{V\pi}$, $\lambda_{V\pi}$, and $\tau_{V\pi}$ 
actually diverge in some approximations when the expansion order $N_\ell \rightarrow \infty$.
Here we show how to circumvent such a problem in multiple ways, recovering the correct transport coefficients of 
second-order fluid dynamics with $14$ dynamical moments.
We also validate our results for the diffusion-shear coupling by comparison to a numerical
solution of the Boltzmann equation for the propagation of sound waves in an ultrarelativistic ideal gas.
\end{abstract}

\maketitle

\section{Introduction}
\label{sec:intro}

Relativistic second-order fluid dynamics has become an essential tool in the 
description of the space-time evolution of high-energy phenomena, ranging from
astrophysical systems like accretion flows \cite{Banyuls.1997}, stellar collapse,
gamma-ray bursts, and relativistic jets \cite{Begelman.1984,Fryer.2004,Marti.2015,Kouveliotou.2012},
to cosmology \cite{Ellis.2012} and relativistic nuclear collisions at 
BNL-RHIC and CERN-LHC \cite{Muronga:2001zk,Csernai:2006zz,Romatschke:2007mq,Heinz:2013th,Bernhard:2019bmu,Auvinen:2020mpc}.
The space-time evolution of such systems and
the interactions among their constituents are characterized not only 
in terms of an equation of state, but also by non-equilibrium 
transport processes.

The conservation equations $\partial_\mu N^\mu = \partial_\mu T^{\mu\nu} = 0$ 
for the particle four-current $N^\mu$ and the energy-momentum tensor $T^{\mu \nu}$ 
provide $1 + 4 = 5$ equations. For ideal fluids, the conservation laws govern the evolution of the 
equilibrium degrees of freedom in $N^\mu$ and $T^{\mu\nu}$, which are 
identified as the particle number density $n$, energy density $e$, 
and fluid four-velocity $u^\mu$, while the pressure is defined through an equation of state, 
$P \equiv P(e,n)$.
For dissipative fluids, the additional $3 + 6 = 9$ degrees of freedom 
contained in $N^\mu$ and $T^{\mu\nu}$ are the bulk viscous pressure $\Pi$, the particle diffusion current 
$V^\mu$, and the shear-stress tensor $\pi^{\mu\nu} $. Together with the equilibrium fields,
these quantities define the so-called \textit{14 dynamical moments approximation} of relativistic fluid dynamics.

At first order in Knudsen number ${\rm Kn}$, defined as the ratio between the particle mean free 
path $\lambda_{\rm mfp}$ and a characteristic macroscopic length scale $L$, 
the dissipative quantities are given by the asymptotic solutions of more general equations 
of motion, in a manner equivalent to the Navier-Stokes equations.
On the other hand, the inverse Reynolds number ${\rm Re}^{-1}$ characterizes the ratio of a 
dissipative to an equilibrium quantity, e.g., $|\Pi / P|$, $|V^\mu / n|$, and $|\pi^{\mu\nu} / P|$.
In the Navier-Stokes limit, the dissipative quantities, 
which are of first order in ${\rm Re}^{-1}$, are algebraically related to 
the thermodynamic forces, which are of first order in ${\rm Kn}$.
The first-order transport coefficients relating them measure different properties of matter, 
such as viscosity, diffusivity, and thermal or electric conductivity. 
These are also found in the well-known transport laws of Newton, Fick, and Ohm.

Starting from the seminal works of M\"uller \cite{Muller.1967} and Israel and Stewart \cite{Israel.1979}, it became 
evident that, in relativistic fluid dynamics, second-order equations are required in order to preserve causality and stability 
\cite{Muller.1967,Israel.1979,Groot.1980,Hiscock.1983,Hiscock.1985,Cercignani_book,Rezzolla.2013}. When the 
irreducible moments are expressed accurately up to second order in ${\rm Kn}$, ${\rm Re}^{-1}$, or their product, new 
cross-coupling transport coefficients emerge in the transport equations. 
A systematic derivation of all transport coefficients is possible using an underlying microscopic theory, e.g., kinetic theory.

In the 1910's, Chapman and Enskog proposed a procedure to derive the equations of fluid dynamics from the 
Boltzmann equation \cite{Chapman_Cowling_Book.1991,Cercignani.1988}. 
While their method is successful at first order, higher-order extensions yield unstable equations, unless the dissipative 
quantities are promoted to dynamical degrees of freedom \cite{Struchtrup.2004}. 
These problems were already recognized by Grad \cite{Grad:1949} in the late 1940's and led 
to a new framework known as the method of moments in nonrelativistic kinetic theory.

Beyond the regime of applicability of relativistic fluid dynamics
(valid for small ${\rm Kn}$ and ${\rm Re}^{-1}$), kinetic theory should be employed 
for the phase-space evolution of the single-particle distribution 
function. Due to the momentum degrees of freedom and the non-linear collision term, 
kinetic theory is computationally more expensive. 
In the early 1950's, Bhatnagar, Gross, and Krook proposed the celebrated BGK 
relaxation-time approximation (RTA) for the nonrelativistic Boltzmann equation \cite{Bhatnagar.1954}.
The RTA paradigm was extended to relativistic kinetic theory, first by Marle \cite{Marle.1969,Cercignani_book} for 
massive particles and then by Anderson and Witting \cite{Anderson:1974a,Cercignani_book} for both 
massive and massless particles. The simplicity of the RTA allows to derive analytical solutions of the relativistic Boltzmann 
equation, e.g., for the Bjorken \cite{Florkowski:2013lya,Florkowski:2014sfa}, 
Gubser \cite{Denicol:2014xca}, and Hubble flows \cite{Bazow.2016}. 
Such solutions have served as benchmarks for testing the 
validity of the equations of second-order fluid dynamics 
\cite{Florkowski:2013lya,Florkowski:2014sfa,Denicol:2014xca,Bazow.2016,Denicol:2018pak,McNelis:2021zji}.
The successful comparison between kinetic theory and fluid dynamics  
relies on the correct implementation of the first- and second-order 
transport coefficients, which is the topic of the present work.

In this paper we re-derive the transport coefficients arising in the Anderson-Witting 
RTA for the linearized collision term \cite{Anderson:1974a}. We adopt 
the method of moments as formulated by Denicol, Niemi, Moln\'ar, and Rischke 
(in the following reluctantly referred to as DNMR) \cite{Denicol:2012cn}, 
as well as the second-order Chapman-Enskog--like method introduced by Jaiswal and others 
\cite{Jaiswal:2013npa,Panda:2020zhr,Panda:2021pvq}. For the DNMR method, we actually study three
different variants, as explained in the following.

In the method of moments, the deviation $\delta f_\bk = f_\bk - f_{0\bk}$ of the single-particle distribution function 
$f_\bk$ from local equilibrium $f_{0\bk}$ is characterized in terms of its irreducible moments 
$\rho^{\mu_1 \cdots \mu_\ell}_r$. In the \textit{standard} DNMR approach, 
$\delta f_\bk$ is expanded in terms of an orthogonal
basis taking into account the irreducible moments $\rho^{\mu_1 \cdots \mu_\ell}_r$ of order $0 \le r \le N_\ell$. 
This expansion becomes complete in the limit 
$N_\ell \rightarrow \infty$, but truncating it at some finite order $N_\ell$ yields an approximation and not an 
exact representation of $\delta f_\bk$. Furthermore, the moments of negative order $r < 0$ are not explicitly included in 
the expansion of $\delta f_\bk$. They are usually constructed in terms of those that are included in this expansion, hence 
introducing an obvious dependence on the truncation order $N_\ell$ that affects the second-order transport coefficients 
explicitly. 

In the simple case of an ultrarelativistic ideal gas, the basis functions can be computed analytically to arbitrary order. 
The coefficients $\gamma^{(\ell)}_{r0}$ introduced in Ref.~\cite{Denicol:2012cn} connecting 
$\rho^{\mu_1 \cdots \mu_\ell}_{-r}$ to $\rho_0^{\mu_1 \cdots \mu_\ell}$ turn out to diverge when 
$N_\ell \rightarrow \infty$. This behavior can be traced back to $O({\rm Kn})$ contributions that are not contained in 
$\gamma^{(\ell)}_{r0}$. Taking the missing contributions explicitly into account following 
Ref.~\cite{Wagner:2022ayd} leads to corrected coefficients $\Gamma^{(\ell)}_{r0}$, which still remain functions 
of $N_\ell$, but are no longer divergent.

As a second approach to compute the transport coefficients within the DNMR framework, we also consider the 
so-called \textit{shifted-basis} approach, i.e., an
expansion of $\delta f_\bk$  where a shift $s_\ell$ is employed for the moments of tensor rank $\ell$. 
This explicitly accounts for moments of order 
$-s_\ell \le r \le N_\ell$ in the expansion of $\delta f_\bk$, such that the representation of the negative-order moments 
with $-s_\ell \le r < 0$ becomes independent of $N_\ell$.

Finally, due to the simple structure of the RTA collision term, 
the negative-order moments can be obtained directly from the 
moment equations, without resorting to basis-dependent representations. 
We refer to this third DNMR-type method as the \textit{basis-free} approach. 

For completeness, we also employ the second-order Chapman-Enskog method introduced in 
Ref.~\cite{Jaiswal:2013npa}. Our results are in agreement with the $N_\ell \rightarrow \infty$ limit of those obtained 
using the method of moments, but differ from those reported in 
Refs.~\cite{Jaiswal:2013npa,Panda:2020zhr,Panda:2021pvq}, obtained using the second-order 
Chapman-Enskog method. We point out that this discrepancy is due to the omission of second-order contributions, 
which we derive explicitly. 

We provide further validation of our results for the RTA by an explicit numerical example focusing on longitudinal waves 
propagating through an ultrarelativistic ideal gas, where the mixing of the shear and diffusion modes is characterized by 
$\ell_{V\pi}$. So far, this second-order transport coefficient was reported as
$\ell_{V\pi} \neq 0$. However, comparing the numerical solution of the Boltzmann equation \cite{Ambrus:2017keg} and 
the results of the second-order fluid-dynamical equations confirms that, in RTA, $\ell_{V\pi} = 0$.

This paper is organized as follows.
We review the method of moments applied to the relativistic Boltzmann equation in Sec.~\ref{sec:BTE}.
In Sec.~\ref{sec:RTA}, we derive the transport coefficients of second-order fluid dynamics using the RTA for the 
collision term.
In Sec.~\ref{sec:UR}, we calculate these transport coefficients for an ultrarelativistic ideal gas and validate our results in 
Sec.~\ref{sec:long} by comparison with the numerical solution of the full Boltzmann equation in RTA 
in the context of the propagation of longitudinal waves. Section~\ref{sec:conc} concludes this paper with a summary of
our results.

In this paper we work in flat space-time with metric tensor 
$g_{\mu \nu}=\text{diag}(1,-1,-1,-1)$, and adopt natural units $\hbar=c=k_{B}=1$. 
The fluid-flow four-velocity $u^{\mu}=\gamma (1,\mathbf{v})$
is time-like and normalized, $u^{\mu}u_{\mu}= 1$, such that $\gamma=(1-\mathbf{v}^{2})^{-1/2}$.
The local rest frame (LRF) of the fluid is defined by $u_{\text{LRF}}^{\mu }=(1,\mathbf{0})$. 
The rank-two projection operator onto the three-space orthogonal to 
$u^{\mu}$ is defined as $\Delta ^{\mu \nu } \equiv g^{\mu \nu}-u^{\mu}u^{\nu}$. 
The symmetric, traceless, and orthogonal
projection tensors of rank $2\ell$, $\Delta_{\nu _{1}\cdots \nu_{\ell }}^{\mu_{1}\cdots\mu _{\ell }}$, are constructed 
using rank-two projection operators. 
The projection of tensors $A^{\mu_1 \cdots \mu_\ell}$ is
denoted as $A^{\left\langle \mu _{1}\cdots \mu_{\ell}\right\rangle } 
\equiv \Delta_{\nu_{1}\cdots \nu_{\ell }}^{\mu_{1}\cdots \mu_{\ell }}A^{\nu_{1}\cdots \nu_{\ell }}$.

The comoving derivative $D\equiv u^{\mu}\partial _{\mu }$ of a quantity $A$ is denoted by 
$\dot{A} = D A \equiv u^{\nu }\partial_{\nu}{A}$, while the gradient operator is denoted by
$\nabla_{\nu}A\equiv \Delta _{\nu}^{\alpha}\partial_{\alpha}A$. 
Therefore, the four-gradient is decomposed as 
$\partial_{\mu}\equiv u_{\mu }D + \nabla_{\mu}$, 
hence $\partial_{\mu} u_{\nu}\equiv u_{\mu}\dot{u}_{\nu} 
+\nabla_{\mu}u_{\nu}=u_{\mu}\dot{u}_{\nu} + \frac{1}{3}\theta \Delta_{\mu \nu }
+\sigma_{\mu\nu}+\omega_{\mu \nu}$, 
where $\theta\equiv \nabla_{\mu}u^{\mu }$ is the expansion scalar,  
$\sigma^{\mu \nu }\equiv\nabla^{\left\langle \mu \right. } 
u^{\left. \nu \right\rangle }=\frac{1}{2}(\nabla^{\mu}u^{\nu }+\nabla^{\nu}u^{\mu })
-\frac{1}{3}\theta \Delta^{\mu \nu}$ is the shear tensor, and 
$\omega^{\mu \nu}\equiv \frac{1}{2}(\nabla^{\mu}u^{\nu}-\nabla^{\nu}u^{\mu})$ is the vorticity.

The four-momentum $k^{\mu}=(k^0,\mathbf{k})$ of particles is normalized to their rest mass
squared, $k^{\mu }k_{\mu }=m_{0}^{2}$, where $k^{0}=\sqrt{\mathbf{k}^{2}+m_{0}^{2}}$ is the on-shell energy of particles.    
We define the energy variable $E_{\mathbf{k}}\equiv k^{\mu }u_{\mu }$ and the projected momentum
$k^{\left\langle \mu \right\rangle } \equiv \Delta _{\nu }^{\mu }k^{\nu }$, such that
$k^{\mu} = E_{\mathbf{k}} u^{\mu} + k^{\langle \mu \rangle}$. In the LRF, 
$E_{\mathbf{k}}= k^0$ is the energy and $k^{\left\langle \mu \right\rangle } = (0,\mathbf{k})$ is the three-momentum.

Integrals over momentum space are abbreviated with angular brackets,
$\left\langle \cdots \right\rangle \equiv \int dK \cdots f_{\mathbf{k}}$,
$\left\langle \cdots \right\rangle _0 \equiv \int dK \cdots f_{0\mathbf{k}}$ 
and $\left\langle \cdots \right\rangle_{\delta }\equiv \int dK\cdots \delta f_{\mathbf{k}}$.
Here, $dK \equiv g d^{3}\mathbf{k}/[(2\pi )^{3}k^{0}]$ is the invariant 
measure in momentum space and $g$ is the degeneracy factor of a momentum state.

\section{Method of moments}
\label{sec:BTE}

In this section, we recall the method of moments introduced in Ref.~\cite{Denicol:2012cn}. 
In Sec.~\ref{sec:BTE:Drho}, the equations of motion for the irreducible moments are presented. 
The expansion of $\delta f_{\bk}$ is discussed in Sec.~\ref{sec:BTE:delta_f_expansion}, extending the 
standard DNMR approach of Ref.~\cite{Denicol:2012cn} to explicitly contain moments with negative indices by 
using a shifted orthogonal basis. The power-counting scheme required to close the system of equations of motion
for the irreducible moments is discussed for the standard approach and the shifted-basis approach 
in Secs.~\ref{sec:BTE:DNMR_approach} and \ref{sec:BTE:shifted}, respectively.

\subsection{Equations of motion for the irreducible moments}\label{sec:BTE:Drho}

The relativistic Boltzmann equation \cite{Groot.1980,Cercignani_book} for the single-particle distribution function 
$f_{\mathbf{k}}$ reads 
\begin{equation}
k^{\mu }\partial_{\mu } f_{\mathbf{k}}=C\left[ f\right]\; ,
\label{BTE}
\end{equation}
where $C[f]$ is the collision term.
Local equilibrium is defined by $C\left[ f_0\right] = 0$, 
which is fulfilled by the J\"uttner distribution \cite{Juttner} 
\begin{equation}
f_{0\mathbf{k}}=\left[ \exp \left( \beta E_{\mathbf{k}}-\alpha\right) +a\right] ^{-1}\;,  \label{f_0k}
\end{equation}
with $\alpha=\mu\beta$, where $\mu$ is the chemical potential
and $\beta=1/T$ the inverse temperature, while $a=\pm 1$ for fermions/bosons 
and $a\rightarrow 0$ for Boltzmann particles. We also introduce the notation 
$\bar{f}_{0\bk} = 1-af_{0\mathbf{k}}$.

In local equilibrium, the particle four-current $N^{\mu}_0 \equiv \left\langle k^\mu \right\rangle_0$ 
and the energy-momentum tensor $T^{\mu \nu}_0 \equiv \left\langle k^\mu k^\nu\right\rangle_0$ of the fluid are 
\begin{align}
N^{\mu}_0 &= n u^{\mu}\;, &
T^{\mu \nu}_0 &= e u^{\mu}u^{\nu} - P \Delta^{\mu\nu}\;.
\end{align}
The tensor projections of these quantities represent the particle density,
energy density, and isotropic pressure, 
\begin{align}
n &\equiv N_{0}^{\mu}u_{\mu}=\left\langle E_{\mathbf{k}} \right\rangle_{0}\;,  \quad
e \equiv T_{0}^{\mu \nu}u_{\mu}u_{\nu}=\left\langle E_{\mathbf{k}}^{2}\right\rangle_{0}\;,  \nonumber\\
P &\equiv  -\frac{1}{3} T_{0}^{\mu \nu }\Delta_{\mu \nu} 
=-\frac{1}{3}\left\langle \Delta_{\mu \nu }k^{\mu }k^{\nu}\right\rangle_{0}\;, \label{P0}
\end{align}
where the pressure is related to energy and particle density through an equation of state, 
$P \equiv P(e,n) = P(\alpha, \beta)$.

The irreducible moments of $\delta f_{\mathbf{k}}$ are defined as
\begin{equation}
\rho _{r}^{\mu _{1}\cdots \mu _{\ell}}\equiv \left\langle 
E_{\mathbf{k}}^{r}k^{\left\langle \mu_{1}\right. }\cdots 
k^{\left. \mu_{\ell}\right\rangle}\right\rangle_{\delta }\;,  
\label{rho_r_general}
\end{equation}
where $r$ denotes the power of energy $E_{\mathbf{k}}$ and
$k^{\left\langle \mu _{1}\right. }\cdots k^{\left. \mu _{\ell}\right\rangle }
=\Delta_{\nu_{1}\cdots \nu_{\ell }}^{\mu_{1}\cdots \mu_{\ell }}k^{\nu_{1}}\cdots k^{\nu_{\ell }}$
are the irreducible tensors forming an orthogonal basis \cite{Denicol:2012cn,Groot.1980}.

The out-of-equilibrium particle four-current and energy-momentum tensor are defined as
\begin{align}
N^{\mu} &\equiv \left\langle k^\mu \right\rangle 
=\left\langle k^\mu \right\rangle_0 + \left\langle k^\mu \right\rangle_{\delta} 
= \left(n + \rho_1 \right) u^{\mu} + V^{\mu}\;, 
\label{N_mu}\\
T^{\mu \nu} &\equiv \left\langle k^\mu k^\nu \right\rangle 
= \left\langle k^\mu k^\nu\right\rangle_0 + \left\langle k^\mu k^\nu \right\rangle_{\delta} 
\nonumber \\
&= \left(e + \rho_2\right) u^{\mu}u^{\nu} 
- \left(P + \Pi \right)\Delta^{\mu\nu} + 2 \rho_1^{(\mu} u^{\nu)} + \pi^{\mu \nu}\;, 
\label{T_munu}
\end{align}
where the particle diffusion four-current and the shear-stress tensor are defined by
\begin{align}
V^{\mu} &\equiv  \Delta^{\mu}_{\alpha} N^{\alpha} =\left\langle k^{\langle \mu \rangle} \right\rangle_{\delta} 
\equiv \rho^{\mu}_0\;, 
\label{V_mu} \\ 
\pi^{\mu \nu} &\equiv  \Delta^{\mu \nu}_{\alpha \beta}T^{\alpha \beta} 
= \left\langle k^{\langle \mu} k^{\nu \rangle}\right\rangle_\delta  \equiv \rho^{\mu \nu}_0\;.  
\label{pi_munu} 
\end{align}
In the Landau frame \cite{Landau_book}, the fluid flow velocity is determined as 
the time-like eigenvector of the energy-momentum tensor, $e u^\mu =T^{\mu \nu} u_{\nu}$, such that
\begin{equation}
\rho^{\mu}_1 \equiv \Delta^{\mu}_{\alpha} T^{\alpha \beta} u_{\beta} 
=\left\langle E_{\bf k} k^{\langle \mu \rangle} \right\rangle_{\delta} = 0 \;.
\label{Landau_flow}
\end{equation}
Furthermore, in order to determine the chemical potential and the temperature, we apply the 
Landau matching conditions \cite{Anderson:1974a},
\begin{align}
\rho_1 &\equiv  \left(N^{\mu } - N^{\mu}_0 \right) u_{\mu }=\left\langle E_{\mathbf{k}} \right\rangle_{\delta} = 0\;,  
\label{delta_n} \\ 
\rho_2 &\equiv \left(T^{\mu \nu } - T_{0}^{\mu \nu } \right)u_{\mu }u_{\nu }
=\left\langle E_{\mathbf{k}}^{2}\right\rangle_{\delta} = 0\;,
\label{delta_e}
\end{align}
such that the bulk viscous pressure can be obtained as
\begin{equation}
\Pi \equiv  -\frac{1}{3} \left(T^{\mu \nu } -  T_{0}^{\mu \nu } \right)\Delta_{\mu \nu } 
=-\frac{1}{3}\left\langle \Delta_{\mu \nu }k^{\mu }k^{\nu }\right\rangle_{\delta}
\equiv -\frac{m^2_0}{3} \rho_0 \;.
\label{Pi}
\end{equation}

The comoving derivative of the irreducible moments, 
$\dot{\rho}_{r}^{\left\langle \mu _{1}\cdots \mu _{\ell }\right\rangle }
\equiv \Delta_{\nu _{1}\cdots \nu _{\ell }}^{\mu _{1}\cdots \mu _{\ell }}
D\rho _{r}^{\nu _{1}\cdots \nu_{\ell }}$, is derived from the
Boltzmann equation (\ref{BTE}), leading to an infinite set of coupled equations of motion.
For the sake of completeness we recall these equations of motion up to rank $2$, 
see Eqs.~(35)--(46) in Ref.~\cite{Denicol:2012cn},
\begin{align}
\dot{\rho}_{r}-C_{r-1}& =\alpha _{r}^{\left( 0\right) }\theta 
+\frac{G_{3r}}{D_{20}}\partial_{\mu }V^{\mu } -\nabla _{\mu }\rho _{r-1}^{\mu }  
+ r\rho _{r-1}^{\mu } \dot{u}_{\mu } \notag \\
&+ \frac{\theta }{3}\left[m_{0}^{2}(r-1)\rho _{r-2}-(r+2)\rho _{r} - 
3\frac{G_{2r}}{D_{20}}\Pi \right]  \notag \\
& +\left[ (r-1)\rho _{r-2}^{\mu \nu }+ \frac{G_{2r}}{D_{20}}\pi^{\mu \nu }\right] 
\sigma_{\mu \nu } \;, \label{D_rho}
\end{align}
\begin{widetext}
\begin{align}
\dot{\rho}_{r}^{\left\langle \mu \right\rangle }-C_{r-1}^{\left\langle \mu
\right\rangle }& =\alpha _{r}^{\left( 1\right) }\nabla ^{\mu }\alpha
+r\rho _{r-1}^{\mu \nu }\dot{u}_{\nu }-
\frac{1}{3}\nabla ^{\mu }\left( m_{0}^{2}\rho _{r-1}-\rho _{r+1}\right)
-\Delta_{\alpha }^{\mu }\left( \nabla _{\nu }\rho _{r-1}^{\alpha \nu}
+\alpha _{r}^{h}\partial _{\kappa }\pi ^{\kappa \alpha }\right)  \notag \\
& +\frac{1}{3}\left[ m_{0}^{2}\left( r-1\right) \rho _{r-2}^{\mu }
-\left(r+3\right) \rho _{r}^{\mu }\right] \theta 
+\frac{1}{5}\sigma ^{\mu \nu }\left[ 2 m_{0}^{2}\left(r-1\right) \rho_{r-2,\nu }
-\left( 2r+3\right) \rho _{r,\nu }\right]\notag \\
& +\frac{1}{3}\left[ m_{0}^{2}r\rho _{r-1}-\left( r+3\right) \rho_{r+1}
-3\alpha _{r}^{h}\Pi \right] \dot{u}^{\mu }+\alpha _{r}^{h}\nabla^{\mu }\Pi 
+\rho_{r,\nu } \omega^{\mu \nu }  
+\left(r-1\right) \rho_{r-2}^{\mu \nu \lambda} \sigma_{\nu \lambda }  \;,  \label{D_rho_mu}
\end{align}
and 
\begin{align}
\dot{\rho}_{r}^{\left\langle \mu \nu \right\rangle }-C_{r-1}^{\left\langle\mu \nu \right\rangle }
&= 2\alpha _{r}^{\left( 2\right) }\sigma ^{\mu \nu } +
\frac{2}{15}\left[ m_{0}^{4}\left( r-1\right) \rho _{r-2}-m_{0}^{2}\left(2r+3\right) \rho _{r}
+\left( r+4\right) \rho _{r+2}\right] \sigma ^{\mu \nu} 
+2\rho_{r}^{\lambda \left\langle \mu \right. }\omega_{\left. {}\right. \lambda }^{\left. \nu \right\rangle }   \notag \\
&+\frac{2}{5}\dot{u}^{\left\langle \mu \right.}
\left[ m_{0}^{2}r\rho_{r-1}^{\left. \nu \right\rangle } 
-\left( r+5\right) \rho _{r+1}^{\left. \nu \right\rangle }\right]  
- \frac{2}{5} \nabla ^{\left\langle \mu \right. } 
\left( m_{0}^{2}\rho_{r-1}^{\left. \nu \right\rangle }
-\rho_{r+1}^{\left. \nu \right\rangle}\right)
+ \frac{1}{3}\left[ m_{0}^{2}\left( r-1\right) \rho _{r-2}^{\mu \nu} 
-\left( r+4\right) \rho_{r}^{\mu \nu }\right] \theta 
\notag \\
&+ \frac{2}{7}\left[2m_{0}^{2}\left( r-1\right) \rho_{r-2}^{\lambda \left\langle \mu \right.}
-\left( 2r+5\right) \rho_{r}^{\lambda \left\langle \mu \right. }\right]
\sigma_{\lambda }^{\left. \nu \right\rangle }
+ r\rho_{r-1}^{\mu \nu \gamma} \dot{u}_{\gamma }
-\Delta_{\alpha \beta }^{\mu \nu }\nabla _{\lambda }\rho _{r-1}^{\alpha \beta\lambda }
+\left( r-1\right) \rho _{r-2}^{\mu \nu \lambda \kappa }\sigma_{\lambda \kappa}\;,
\label{D_rho_munu}
\end{align}
\end{widetext}
where the irreducible moments of the collision term are
\begin{equation}
C_{r-1}^{\left\langle \mu _{1}\cdots \mu _{\ell }\right\rangle }
= \int dK\,E_{\mathbf{k}}^{r-1}\,k^{\langle \mu_1} \cdots k^{\mu_\ell \rangle} C\left[ f\right]\; .  
\label{eq:C_r}
\end{equation}
In the above, $\alpha_r^h = -\beta J_{r+2,1}/(n h)$, where the enthalpy per particle is $h\equiv \left( e+P\right) /n$, while 
\begin{align}
\alpha_r^{(0)} &= (1 - r) I_{r1} - I_{r0} - \frac{n}{D_{20}} 
\left(h G_{2r} - G_{3r} \right)\;,\label{alpha0}\\
\alpha_r^{(1)} &= J_{r+1,1} - \frac{J_{r+2,1}}{h} \;,\label{alpha1}\\
\alpha_r^{(2)} &= I_{r+2,1} + (r - 1) I_{r+2,2}\;.\label{alpha2}
\end{align}
The primary and auxiliary thermodynamic integrals, $I_{nq}(\alpha, \beta)$ and $J_{nq}(\alpha, \beta)$, respectively, 
are defined as
\begin{align} \label{I_nq}
I_{nq} &= \frac{(-1)^q}{(2q+1)!!} \braket{E_{\bk}^{n-2q} (\Delta^{\alpha\beta}k_\alpha k_\beta)^q}_0\;, 
\\ \label{J_nq}
J_{nq} &\equiv \left. \frac{\partial I_{nq}}{\partial \alpha}\right|_{\beta} 
= \beta^{-1} \left[ I_{n-1,q-1} + \left(n-2q\right) I_{n-1,q} \right] \;. 
\end{align}
Furthermore, in the above equations, we also introduced the functions
\begin{align}
G_{nm} &= J_{n0} J_{m0} - J_{n-1,0} J_{m+1,0}\;, \\
D_{nq} &= J_{n+1,q} J_{n-1,q} - J_{nq}^2\;.
\label{eq:Idef}
\end{align}

The conservation of particle number $\partial_\mu N^{\mu} = 0$, energy $u_\nu\partial_\mu T^{\mu \nu} =0$, 
and momentum $\Delta^{\mu}_{\beta} \partial_\alpha T^{\alpha \beta} =0$ can be written in the form
\begin{align} \label{eq:cons_n}
	\dot{n} + n \theta + \partial_\mu V^\mu &= 0\;,\\  \label{eq:cons_e}
	\dot{e} + (e + P + \Pi) \theta - \pi^{\mu\nu} \sigma_{\mu\nu} &= 0\;, \\ \label{eq:cons_u}
(e + P + \Pi) \dot{u}^\mu - \nabla^\mu(P + \Pi) 
+ \Delta^\mu{}_\lambda \partial_\nu \pi^{\lambda\nu} &= 0\;. 
\end{align}
In order to solve these equations, we have to provide equations of motion for the dissipative quantities 
$\Pi$, $V^\mu$, and $\pi^{\mu \nu}$. In the next sections, we will show
how to obtain them from Eqs.~(\ref{D_rho})--(\ref{D_rho_munu}) based on different series expansions 
and approximations.

\subsection{Expansion of the distribution function in momentum space}
\label{sec:BTE:delta_f_expansion}

The equations of motion for the primary dissipative quantities $\rho_{0} = -3 \Pi/m^2_0$, 
$\rho_{0}^{\mu} = V^{\mu}$, and $\rho_{0}^{\mu \nu} = \pi^{\mu\nu}$ also include negative-order moments 
$\rho^{\mu_1 \cdots \mu_\ell}_{r < 0}$. From the right-hand sides of 
Eqs.~\eqref{D_rho}--\eqref{D_rho_munu} (for $r=0$) we observe that these are
\begin{equation}
	\rho_{-2}\;, \rho_{-1}\;, \rho_{-2}^\mu\;, \rho_{-1}^\mu\;, 
	\rho_{-2}^{\mu\nu}\;, \rho_{-1}^{\mu\nu}\;. \label{needed_rhoneg_m}
\end{equation}
Note that these equations formally also involve the moments $\rho_1, \rho_2$, and $\rho_1^\mu$, which, however,
vanish due to the Landau matching conditions and the choice of the Landau frame for the fluid velocity.
Furthermore, there are tensors of rank $\ell > 2$. 
These are omitted in the following, since they are of higher order in Knudsen and inverse Reynolds number, 
$\rho^{\mu\nu \lambda \cdots}_r \simeq O(\Kn^2, \Re \Kn)$, see Ref.~\cite{Denicol:2012cn} for a discussion.

Following the suggestions of Refs.~\cite{Thorne_1981,Struchtrup_1998,Eu_book} we 
consider the expansion of $\delta f_{\mathbf{k}} = f_\bk - f_{0\mathbf{k}}$ with respect to 
a complete and orthogonal basis,
\begin{equation}
	\delta f_{\mathbf{k}} = f_{0\mathbf{k}} \bar{f}_{0\bk}
	\sum_{\ell =0}^{\infty }\sum_{n=0}^{N_{\ell } + s_\ell} 
	\rho_{n-s_\ell}^{\mu_{1}\cdots \mu_{\ell }}
	E_\bk^{-s_\ell}
	k_{\left\langle \mu _{1}\right. }\cdots k_{\left.\mu _{\ell }\right\rangle} 
	\widetilde{\mathcal{H}}_{\mathbf{k}n}^{(\ell)}\;,	
	\label{f_k_expansion}
\end{equation}
where the factor $E_\bk^{-s_\ell}$ allows the expansion to contain 
moments with negative energy index, hence naturally accounting for all moments 
$\rho_r^{\mu_{1}\cdots \mu_{\ell }}$ with $-s_\ell \le r \le N_\ell$.
In general, $N_\ell$ and the shift $s_\ell$ can be set to different values for each tensor rank $\ell$. 

We note that Eq.~\eqref{f_k_expansion} generalizes the expansion of Ref.~\cite{Denicol:2012cn},  
recovering it when $s_\ell = 0$.
In the above and in what follows, we use an overhead tilde $\widetilde{\phantom{a}}$ to denote quantities which 
differ from the ones introduced in Ref.~\cite{Denicol:2012cn}. 
When discussing the $s_\ell = 0$ case, all overhead tildes will be dropped, $
\widetilde{A} \xrightarrow[]{s_\ell = 0} A$.

The coefficient $\widetilde{\mathcal{H}}_{\mathbf{k}n}^{(\ell )}$ is a polynomial in
energy of order $N_\ell + s_\ell$,
\begin{equation}
\widetilde{\mathcal{H}}_{\mathbf{k}n}^{(\ell )} =\frac{(-1)^{\ell}}{\ell! J_{2\ell-2s_\ell,\ell}} 
\sum_{m=n}^{N_\ell+s_\ell} \widetilde{a}_{mn}^{(\ell)}
\widetilde{P}_{\mathbf{k} m}^{(\ell)}\;, \label{eq:Hfunction_k}
\end{equation}
where 
\begin{equation}
\widetilde{P}_{\mathbf{k} m}^{(\ell)} = \sum_{r=0}^{m}
\widetilde{a}_{mr}^{(\ell)}E_{\mathbf{k}}^{r}
\label{eq:P_k}
\end{equation}
is a polynomial of order $m$ in energy.
The $\widetilde{a}_{mn}^{(\ell)}$ coefficients are obtained through the Gram-Schmidt procedure imposing the following orthogonality condition:
\begin{equation}
	\int dK \widetilde{\omega}^{(\ell)} \widetilde{P}_{\bk m}^{(\ell)} 
	\widetilde{P}_{\bk n}^{(\ell)} = \delta_{mn}\;,
	\label{eq:P_ortho}
\end{equation}
where the weight $\widetilde{\omega}^{(\ell)}$ is defined as
\begin{equation}
	\widetilde{\omega}^{(\ell)} = \frac{(-1)^{\ell}}{(2\ell + 1)!!} \frac{ E_\bk^{-2s_\ell}}{J_{2\ell-2s_\ell,\ell} } 
	(\Delta^{\alpha\beta} k_\alpha k_\beta)^\ell f_{0\mathbf{k}} \bar{f}_{0\bk}\;.
	\label{eq:omega_def}
\end{equation}

If $N_\ell \rightarrow \infty$, the expansion (\ref{f_k_expansion}) is exact. A finite
$N_\ell + s_\ell < \infty$ defines a truncation, i.e., the set of irreducible moments
$\rho_r^{\mu_1 \cdots \mu_\ell}$, $-s_\ell \le r \le N_\ell$ used to approximate $\delta f_\bk$.
Consequently, we must be able to recover any $\rho_r^{\mu_1 \cdots \mu_\ell}$ contained in this set from 
this particular truncation of $\delta f_\bk$. In order to see this, we define the function
\begin{align}
	\widetilde{\mathcal{F}}_{\mp rn}^{\left( \ell \right) } =& (-1)^\ell \ell !
	J_{2\ell - 2s_\ell,\ell} 
	\int dK \, \widetilde{\omega}^{(\ell)} E_{\mathbf{k}}^{\pm r} \widetilde{\mathcal{H}}_{\mathbf{k}n}^{\left( \ell \right)}
	\nonumber\\
	=& \sum_{m = n}^{N_\ell + s_\ell} \sum_{q = 0}^m  
 \frac{J_{\pm r + q + 2\ell - 2s_\ell, \ell}}{J_{2\ell - 2s_\ell,\ell}} \widetilde{a}^{(\ell)}_{mn} \widetilde{a}^{(\ell)}_{mq}\;.
	\label{F_rn}
\end{align}
Then, using Eqs.~(\ref{rho_r_general}) and 
(\ref{f_k_expansion}), any irreducible moment with tensor-rank $\ell$ and of arbitrary order $r$
can be expressed as a linear combination of the rank-$\ell$ moments appearing in 
the expansion \eqref{f_k_expansion}:
\begin{multline}
	\rho_{\pm r - s_\ell}^{\mu _{1}\cdots \mu _{\ell }}
	\equiv \sum_{n=0}^{N_{\ell }+s_\ell}\rho_{n-s_\ell}^{\mu_{1}\cdots \mu_{\ell }}
	\widetilde{\mathcal{F}}_{\mp r,n}^{\left( \ell \right) }  \\
	= \sum_{n=-s_\ell}^{-1}\rho_{n}^{\mu_{1}\cdots \mu_{\ell }}
	\widetilde{\mathcal{F}}_{\mp r,n+s_\ell}^{\left( \ell \right) }  +
	\sum_{n=0}^{N_{\ell }}\rho_{n}^{\mu_{1}\cdots \mu_{\ell }}
	\widetilde{\mathcal{F}}_{\mp r,n+s_\ell}^{\left( \ell \right) }\; .
	\label{useful}
\end{multline}
For indices satisfying $0 \leq i,j\leq N_\ell + s_\ell$, we have 
$\widetilde{\mathcal{F}}_{-i,j}^{\left( \ell \right) }=\delta_{ij}$ by construction, 
hence Eq.~\eqref{useful} reduces to an identity.
On the other hand, for any $r > 0$, 
the moments $\rho_{-r - s_\ell}^{\mu_{1}\cdots \mu_{\ell }}$ and 
$\rho_{N_\ell + r}^{\mu_1 \cdots \mu_\ell}$, which are not contained in the expansion \eqref{f_k_expansion},
can be expressed in terms of a sum over those moments which do appear in Eq.~\eqref{f_k_expansion}.

The shifts $s_\ell$ introduced in Eq.~\eqref{f_k_expansion} are in principle arbitrary. However, note that in the 
massless case infrared divergences can appear due to negative powers of energy $E_\bk^{-s_\ell}$. 
In order to avoid these, the maximum possible value of the shift is given by
\begin{equation}
	s^{\rm max}_\ell = \ell\;, \text{ when $m_0 = 0$}\;.
	\label{shift_m0}
\end{equation}
This corresponds to the orthogonal basis
$1$, $v^{\left\langle \mu_{1}\right\rangle}$, $v^{\left\langle \mu_{1}\right.} v^{\left.\mu_{2}\right\rangle}$, \dots, 
$v^{\left\langle \mu_{1}\right. }\cdots v^{\left.\mu_{\ell }\right\rangle}$ 
of Ref.~\cite{Eu_book}, where
\begin{equation}
	v^{\left\langle \mu \right\rangle } \equiv 
	\frac{k^{\left\langle \mu \right\rangle }}{  E_{\mathbf{k}}} 
	= \frac{k^{\mu}}{E_{\mathbf{k}}} - u^{\mu}\;, 
\end{equation}
while the generalization to rank-$\ell$ tensors reads
$v^{\left\langle \mu_{1}\right. }\cdots v^{\left.\mu _{\ell }\right\rangle} = 
E^{-\ell}_{\mathbf{k}} k^{\left\langle \mu_{1}\right. }\cdots k^{\left.\mu_{\ell }\right\rangle}$.
This velocity-based orthogonal basis is also convenient for calculating the nonrelativistic limits of the moments 
\cite{Eu_book}.

Finally, in the case of finite particle mass, the negative-order moments appearing in Eq.~\eqref{needed_rhoneg_m} 
can be included in Eq.~\eqref{f_k_expansion} using the following parameters, 
\begin{equation}
	s_0 = s_1 = s_2 = 2\;, \text{ when $m_0 > 0$}\,. \label{shift_m}
\end{equation}
\subsection{Power counting in the standard DNMR approach}
\label{sec:BTE:DNMR_approach}

One can show~\cite{Denicol:2012cn} that in the case of binary collisions the
linearized collision integral reads
\begin{equation}
	C_{r-1}^{\left\langle \mu _{1}\cdots \mu _{\ell }\right\rangle }
	= -\sum_{n=0}^{N_{\ell } + s_\ell} \mathcal{A}_{r,n-s_\ell}^{\left( \ell \right)} \rho_{n-s_\ell}^{\mu_{1}\cdots \mu_{\ell}}\;,
	\label{Lin_collint}
\end{equation}
where $-s_\ell \le r \le N_\ell$. 
In the above, $\mathcal{A}_{rn}^{\left( \ell \right) }\sim \lambda _{\mathrm{mfp}}^{-1}$ 
is the collision matrix while its inverse $\tau_{rn}^{\left( \ell \right) } = \left(\mathcal{A}^{(\ell)} \right)_{rn}^{-1}$ 
is related to microscopic time scales proportional to the mean free time between collisions.

This introduces a natural power-counting scheme in terms of ${\rm Kn}$ and ${\rm Re}^{-1}$, 
allowing second-order fluid dynamics to be derived systematically from the equations of motion for the irreducible moments. 
In particular, we will apply this power-counting scheme also to the negative-order moments.

As stated before, the equations of motion for the dissipative quantities follow from  
Eqs.~(\ref{D_rho})--(\ref{D_rho_munu}) by choosing  $r=0$, i.e., 
the lowest-order irreducible moments appearing in Eqs.~(\ref{N_mu})--(\ref{T_munu}).
In this way, these moments are chosen to be dynamical, i.e., they represent the solution of the corresponding partial differential equations.
However, since we are dealing with an infinite hierarchy of moment equations, we are also obliged to 
determine the remaining moments with $r\neq 0$. 

Following Ref.~\cite{Denicol:2012cn} the moment equations for 
$0 < r \leq N_\ell$ are approximated by their asymptotic solutions as 
\begin{align}
	\rho_{r> 0} &\simeq -\frac{3}{m_0^2}\Omega^{(0)}_{r0} \Pi 
	+\frac{3}{m_0^2} (\zeta_r - \Omega^{(0)}_{r0} \zeta_0) \theta\;, 
	\label{eq:DNMR_matching_rho} \\
	\rho^\mu_{r> 0} &\simeq \Omega^{(1)}_{r0} V^\mu
	+ (\kappa_r - \Omega^{(1)}_{r0} \kappa_0) \nabla^\mu \alpha\; , 
	\label{eq:DNMR_matching_rho_mu} \\
	\rho^{\mu\nu}_{r> 0} &\simeq \Omega^{(2)}_{r0} \pi^{\mu\nu} 
	+ 2(\eta_r - \Omega^{(2)}_{r0} \eta_0) \sigma^{\mu\nu}\; ,
	\label{eq:DNMR_matching_rho_munu}
\end{align}
where the first-order transport coefficients $\zeta_r$, $\kappa_r$, and $\eta_r$ are
\begin{gather} 
	\zeta_r \equiv \frac{m_0^2}{3} \sum_{n = 0, \neq 1,2}^{N_0} \!
	\tau_{rn}^{(0)} \alpha^{(0)}_n\;, \nonumber \\ 
	\kappa_r \equiv \sum_{n = 0, \neq 1}^{N_1} \!
	\tau^{(1)}_{rn} \alpha^{(1)}_n\;, 
	\label{NS_coeff}
	\quad 
	\eta_r \equiv \sum_{n = 0}^{N_2} 
	\tau^{(2)}_{rn} \alpha^{(2)}_n\;.
\end{gather}
Here, $\Omega^{(\ell)}_{rn}$ diagonalizes the collision matrix 
$\mathcal{A}_{rn}^{\left( \ell \right) }$ via $(\Omega^{(\ell)})^{-1} \mathcal{A}^{(\ell)} 
\Omega^{(\ell)} = {\rm diag}(\chi_0^{(\ell)}, 
\chi_1^{(\ell)}, \dots, \chi_{N_\ell}^{(\ell)})$, where without loss of generality the eigenvalues are ordered as 
$\chi^{(\ell)}_0 \le \dots \le \chi^{(\ell)}_{N_\ell}$ and $\Omega^{(\ell)}_{00} = 1$ by convention.

We would like to point out that in the calculations of Refs.~\cite{Denicol:2012cn,Molnar.2014} expressions for the 
moments of negative order $\rho^{\mu_1 \dots \mu_\ell}_{-r}$ were used which neglect terms of order
$O(\textrm{Kn})$. These are obtained by substituting only the 
first terms from the right-hand sides of Eqs.~\eqref{eq:DNMR_matching_rho}--\eqref{eq:DNMR_matching_rho_munu} 
into Eq.~\eqref{useful}, leading to
\begin{gather}
	\rho_{-r} \simeq -\frac{3}{m_0^2}\gamma^{(0)}_{r0} \Pi + O({\rm Kn})\;, \nonumber\\
	\rho_{-r}^\mu \simeq \gamma^{(1)}_{r0} V^\mu + O({\rm Kn})\;, \, \, \,
	\rho_{-r}^{\mu\nu} \simeq \gamma^{(2)}_{r0} \pi^{\mu\nu} + O({\rm Kn})\;, 
	\label{DNMR_rho_neg}
\end{gather}
where the coefficients are
\begin{gather}
\gamma_{r0}^{(0)} = \sum_{n = 0,\neq 1,2}^{N_0} \! \mathcal{F}^{(0)}_{rn}\Omega^{(0)}_{n0}\;,
\nonumber\\
\gamma_{r0}^{(1)} = \sum_{n = 0,\neq 1}^{N_1} \!\mathcal{F}^{(1)}_{rn} \Omega^{(1)}_{n0}\;, \quad 
\gamma_{r0}^{(2)} = \sum_{n = 0}^{N_2} \mathcal{F}^{(2)}_{rn} \Omega^{(2)}_{n0}\;.
\label{gamma}
\end{gather}
However, the neglected $O({\rm Kn})$ contributions to Eq.~\eqref{DNMR_rho_neg} explicitly affect the results for 
the transport coefficients. For instance, in Sec.~\ref{sec:UR}, we show by an explicit calculation that, 
in the case of an ultrarelativistic ideal gas in the RTA, all
$\gamma^{(\ell)}_{r0}$ coefficients actually diverge when $N_\ell \rightarrow \infty$. On the other hand,
taking the $O({\rm Kn})$ contributions into account as described below, 
the modified coefficients will remain finite in this limit.

In order to account for the neglected $O({\rm Kn})$ terms, one first substitutes all terms from 
Eqs.~\eqref{eq:DNMR_matching_rho}--\eqref{eq:DNMR_matching_rho_munu} into Eq.~\eqref{useful}, 
see Ref.~\cite{Wagner:2022ayd}. Then, one replaces the thermodynamic forces using the 
Navier-Stokes relations  $\theta = -\Pi / \zeta_0$, 
$\nabla^\mu \alpha = V^\mu / \kappa_0$, and $\sigma^{\mu\nu} = \pi^{\mu\nu} / (2 \eta_0)$. 
We note that this replacement is a matter of choice. If we did not do this and just kept the
terms as they appear, we would obtain corrections to the
transport coefficients of the $O(\textrm{Kn}^2)$ terms computed in Ref.~\cite{Molnar.2014}, while the
other transport coefficients would not change as compared to their DNMR values. However, in Sec.~\ref{sec:long}
we will see by comparison to the numerical solution of the Boltzmann equation in RTA that the approach described above
leads to a better agreement with the latter, which justifies this procedure.
Ultimately, this leads to a cancelation of the first and third terms on the right-hand sides of 
Eqs.~\eqref{eq:DNMR_matching_rho}--\eqref{eq:DNMR_matching_rho_munu}, such that
\begin{align}
	\rho_{-r} &\simeq -\frac{3}{m_0^2}\Gamma^{(0)}_{r0} \Pi \;, &
	\rho_{-r}^\mu &\simeq \Gamma^{(1)}_{r0} V^\mu\;, &
	\rho_{-r}^{\mu\nu} &\simeq \Gamma^{(2)}_{r0} \pi^{\mu\nu}\;,
	\label{IReD_rhoneg}
\end{align}
where the corrected DNMR coefficients are
\begin{gather}
	\Gamma^{(0)}_{r0} \equiv \sum_{n = 0, \neq 1, 2}^{N_0} \mathcal{F}^{(0)}_{rn} \frac{\zeta_n}{\zeta_0}\;,\nonumber\\ 
	\Gamma^{(1)}_{r0} \equiv \sum_{n = 0, \neq 1}^{N_1} \mathcal{F}^{(1)}_{rn} \frac{\kappa_n}{\kappa_0}\;,
	\quad 
	\Gamma^{(2)}_{r0} \equiv \sum_{n = 0}^{N_2} \mathcal{F}^{(2)}_{rn} \frac{\eta_n}{\eta_0}\;.
	\label{Gamma}
\end{gather}

Recently a different approximation was suggested in Ref.~\cite{Wagner:2022ayd}, called 
Inverse Reynolds Dominance (IReD). 
This is based on a power counting without the diagonalization procedure, 
i.e., without involving Eqs.~\eqref{eq:DNMR_matching_rho}--\eqref{eq:DNMR_matching_rho_munu} as an intermediate step,
but explicitly assuming that 
the non-dynamical moments are approximated by
\begin{align}
	\rho_{r> 0} &\simeq -\frac{3}{m_0^2} \frac{\zeta_r}{\zeta_0} \Pi\;, &
	\rho^\mu_{r> 0} &\simeq \frac{\kappa_r}{\kappa_0} V^\mu\;, &
	\rho^{\mu\nu}_{r> 0} &\simeq \frac{\eta_r}{\eta_0} \pi^{\mu\nu}\;.
	\label{IReD_matching}
\end{align}
Substituting these approximated values into Eq.~\eqref{useful} also leads to the corrected DNMR 
results of Eqs.~\eqref{IReD_rhoneg}--\eqref{Gamma}.
Note that similar approaches made in nonrelativistic \cite{Struchtrup.2004} as well as in multicomponent 
relativistic fluid dynamics \cite{Fotakis:2022usk} are known as the order-of-magnitude approximation.

Comparing Eqs.~(\ref{DNMR_rho_neg})--\eqref{gamma} to Eqs.~(\ref{IReD_rhoneg})--\eqref{Gamma}, 
it becomes clear that moments with 
negative order explicitly depend on the value of the corresponding coefficients, i.e., 
$\gamma_{r0}^{(\ell)}$ or $\Gamma_{r0}^{(\ell)}$.
These approaches lead to transport coefficients that explicitly depend on the truncation order $N_\ell$, while only the latter (corrected) approach
achieves convergence when $N_\ell \rightarrow \infty$. 
In other words, the correct representation of the 
negative-order moments relies on an expansion that includes an 
infinite number of positive-order moments.

\subsection{Power counting in the shifted-basis approach}\label{sec:BTE:shifted}

Employing now the shifted-basis approach to explicitly include negative-order moments in the expansion of 
$\delta f_\bk$, as discussed in Sec.~\ref{sec:BTE:delta_f_expansion}, the relations~\eqref{IReD_matching} 
are generalized in a straightforward manner to 
\begin{gather}
	\rho_{r\ge -s_0} \simeq -\frac{3\zeta_r}{m_0^2 \zeta_0} \Pi\;, \nonumber \\ 
	\rho^\mu_{r \ge -s_1} \simeq \frac{\kappa_r}{\kappa_0} V^\mu\;, \quad
	\rho^{\mu\nu}_{r \ge -s_2} \simeq \frac{\eta_r}{\eta_0} \pi^{\mu\nu}\;.
	\label{shifted_matching}
\end{gather}
The first-order transport coefficients in Eq.~\eqref{NS_coeff} now involve summations also over negative indices:
\begin{gather} 
	\zeta_{r\ge -s_0} \equiv \frac{m_0^2}{3} \sum_{n = -s_0, \neq 1,2}^{N_0} \!
	\tau_{rn}^{(0)} \alpha^{(0)}_n\;, \nonumber\\ 
	\kappa_{r\ge -s_1} \equiv \hspace{-2pt} \sum_{n = -s_1, \neq 1}^{N_1} \hspace{-2pt} 
	\tau^{(1)}_{rn} \alpha^{(1)}_n\;, 
	\ \ 
	\eta_{r \ge -s_2} \hspace{-2pt} \equiv \sum_{n = -s_2}^{N_2} 
	\hspace{-2pt} \tau^{(2)}_{rn} \alpha^{(2)}_n\;.
	\label{shifted_NS_coeff}
\end{gather}

On the other hand, for any finite shift $s_\ell < \infty$, there are always negative-order moments that cannot be 
accounted for in the expansion \eqref{f_k_expansion}. These moments can be 
computed as follows. For $r > 0$, Eq.~\eqref{IReD_rhoneg} can be generalized to yield
\begin{gather}
\rho_{-r-s_0} \simeq -\frac{3}{m_0^2} \widetilde{\Gamma}_{r0}^{(0)} \Pi\;, \nonumber\\
\rho^\mu_{-r-s_1} \simeq \widetilde{\Gamma}_{r0}^{(1)} V^\mu\;, \qquad
\rho^{\mu\nu}_{-r-s_2} \simeq \widetilde{\Gamma}_{r0}^{(2)} \pi^{\mu\nu}\;, \label{shifted_rho_neg} 
\end{gather}
where 
\begin{gather}
\widetilde{\Gamma}_{r0}^{(0)} \equiv \sum_{n = -s_0,\neq 1,2}^{N_0} \! \widetilde{\mathcal{F}}^{(0)}_{r,n+s_0} 
\frac{\zeta_n}{\zeta_0}\;,\nonumber\\
\widetilde{\Gamma}_{r0}^{(1)} \equiv \hspace{-5pt} \sum_{n = -s_1,\neq 1}^{N_1} \hspace{-5pt} 
\widetilde{\mathcal{F}}^{(1)}_{r,n+s_1} \frac{\kappa_n}{\kappa_0}\;, \quad
\widetilde{\Gamma}_{r0}^{(2)} \equiv \!\!\sum_{n = -s_2}^{N_2}\!\! \widetilde{\mathcal{F}}^{(2)}_{r,n+s_2} 
\frac{\eta_n}{\eta_0}\;.
\label{shifted_Gamma}
\end{gather}

As discussed in Eq.~\eqref{shift_m}, setting $s_\ell = 2$ allows the negative-order moments in 
Eq.~\eqref{needed_rhoneg_m} to be expressed using Eq.~\eqref{shifted_matching}, without employing any 
$N_\ell$-dependent $\widetilde{\Gamma}^{(\ell)}_{r0}$ coefficients, 
however an explicit $N_\ell$ dependence still remains at 
the level of the first-order transport coefficients in their definitions, Eq.~\eqref{shifted_NS_coeff}.
As it will become clear in the next section, the transport coefficients obtained using the shifted-basis approach
will become independent of the truncation order in the RTA.

\section{Transient fluid dynamics in the relaxation-time approximation}
\label{sec:RTA}

We begin this section by discussing the Anderson-Witting RTA in Sec.~\ref{sec:RTA:AW}. 
The representation of negative-order moments in the basis-free and shifted-basis approaches are
presented in Secs.~\ref{sec:RTA:rhoneg} and \ref{sec:sbanom}, respectively, 
while the Chapman-Enskog method is employed in Sec.~\ref{sec:RTA:CE}. The second-order transport coefficients for a 
neutral fluid and the additional coefficients appearing in magnetohydrodynamics of charged, 
but unpolarizable fluids are reported in Sec.~\ref{sec:RTA:tcoeffs} and Sec.~\ref{sec:RTA:MHD}, respectively.

\subsection{The Anderson-Witting RTA}
\label{sec:RTA:AW}

The Anderson-Witting RTA for the collision integral reads \cite{Anderson:1974a,Groot.1980,Cercignani_book}
\begin{equation}
	C[f] \equiv -\frac{E_\bk}{\tau_R} (f_\bk - \feq) = -\frac{E_\bk}{\tau_R} \delta f_\bk\;,
	\label{AW}
\end{equation}
where the relaxation time $\tau_R \equiv \tau_R(x^{\mu})$ is a momentum-independent
parameter proportional to the mean free time between collisions.
Substituting the above expression into Eq.~(\ref{eq:C_r}) leads to
\begin{equation}
	C_{r-1}^{\langle \mu_1 \cdots \mu_\ell \rangle} = -\frac{1}{\tau_R}
	\rho_r^{\mu_1 \cdots \mu_\ell}\;. \label{RTA_Cr}
\end{equation}
The matrices $\mathcal{A}^{(\ell)}_{rn}$, $\tau^{(\ell)}_{rn}$, and $\Omega^{(\ell)}_{rn}$ corresponding to the 
collision term \eqref{RTA_Cr} are diagonal,\footnote{The columns of $\Omega^{(\ell)}_{rn}$ can be permuted arbitrarily, 
since all of the eigenvalues $\chi^{(\ell)}_r$ of the collision matrix $\mathcal{A}^{(\ell)}_{rn}$ are equal to $\tau_R$. 
For the sake of simplicity, we choose $\Omega^{(\ell)}_{rn}$ to be diagonal.}
\begin{align}
	\mathcal{A}_{rn}^{\left( \ell \right)} &= \frac{\delta_{rn}}{\tau_R}\;, &
	\tau_{rn}^{\left( \ell \right)} &= \tau_R \delta_{rn}\;, &
	\Omega_{rn}^{(\ell)} &= \delta_{rn}\;.
	\label{RTA_A_tau}
\end{align}
Using these results in Eqs.~\eqref{D_rho}--\eqref{D_rho_munu} and multiplying 
both sides by $\tau_R$ gives 
\begin{align}
	\tau_R \dot{\rho}_{r} + \rho_r 
	& = \tau_R \alpha_{r}^{\left( 0\right) }\theta 
	+ O({\rm Re}^{-1} {\rm Kn})\;, \label{D_rho_RTA} \\  
	\tau_R \dot{\rho}_{r}^{\left\langle \mu \right\rangle } 
	+ \rho_r^{\left\langle \mu\right\rangle } 
	&= \tau_R \alpha_{r}^{\left( 1\right) }\nabla^{\mu }\alpha + O({\rm Re}^{-1} {\rm Kn})\;, \label{D_rho_mu_RTA} \\  
	\tau_R \dot{\rho}_{r}^{\left\langle \mu \nu \right\rangle } 
	+ \rho_r^{\left\langle \mu \nu \right\rangle } 
	&= 2\tau_R \alpha_{r}^{\left( 2\right) }\sigma^{\mu \nu} + O({\rm Re}^{-1} {\rm Kn})\;,
	\label{D_rho_munu_RTA}
\end{align}
where the higher-order terms on the right-hand sides of Eqs.~\eqref{D_rho}--\eqref{D_rho_munu}
were abbreviated by $O({\rm Re}^{-1} {\rm Kn})$ for the sake of simplicity. This implies that all
irreducible moments in these terms are considered to be of order $O(\Re)$, in accordance with our previous discussion.

We also point out that in the RTA all irreducible moments have the same relaxation time, $\tau_R$, 
and hence there is no natural ordering of the eigenvalues $\chi^{(\ell)}_r$ of the collision operator, e.g., see 
Sec.~\ref{sec:BTE:DNMR_approach}. 
Even so, since $\tau_R$ is of first order with respect to ${\rm Kn}$, 
the second-order equations of motion for $\Pi$, $V^\mu$ and $\pi^{\mu\nu}$ 
can still be obtained by replacing all moments $\rho_{r \neq 0}^{\mu_1 \cdots \mu_\ell}$ 
by their first-order approximations, as discussed in 
Secs.~\ref{sec:BTE:DNMR_approach} and \ref{sec:BTE:shifted}. 

The first-order transport coefficients from Eq.~\eqref{NS_coeff} are
\begin{align}
	\zeta_r &= \tau_R \frac{m_0^2}{3} \alpha^{(0)}_r\; , &
	\kappa_r &= \tau_R \alpha^{(1)}_r\; , &
	\eta_r &= \tau_R \alpha^{(2)}_r\; .
	\label{main_RTA_coeffs}
\end{align}
The DNMR coefficients  \eqref{gamma} for the negative-order moments reduce to 
\begin{equation}
	\gamma^{(\ell)}_{r0} = \mathcal{F}^{(\ell)}_{r0}\;.
	\label{RTA_gamma}
\end{equation}
The coefficients \eqref{shifted_Gamma} introduced in the shifted-basis approach are
\begin{gather} 
	\widetilde{\Gamma}^{(0)}_{r0} = \sum_{n = -s_0, \neq 1, 2}^{N_0} \widetilde{\mathcal{F}}^{(0)}_{r,n+s_0} 
	\mathcal{R}^{(0)}_{n0}\; ,\nonumber\\ 
	\widetilde{\Gamma}^{(1)}_{r0} = \hspace{-5pt}\sum_{n = -s_1, \neq 1}^{N_1} \hspace{-5pt} 
	\widetilde{\mathcal{F}}^{(1)}_{r,n+s_1} \mathcal{R}^{(1)}_{n0}\;, 
	\quad
	\widetilde{\Gamma}^{(2)}_{r0} = \!\!\sum_{n = -s_2}^{N_2}\!\! \widetilde{\mathcal{F}}^{(2)}_{r,n+s_2} 
	\mathcal{R}^{(2)}_{n0}\;,
	\label{RTA_Gamma}
\end{gather}
where we introduced
\begin{equation}
	\mathcal{R}^{(\ell)}_{r0} \equiv \frac{\alpha^{(\ell)}_r}{\alpha^{(\ell)}_0}\;.
	\label{R_def}
\end{equation}
The corrected DNMR coefficients corresponding to Eq.~\eqref{Gamma} are obtained by 
setting $s_\ell = 0$ in Eq.~\eqref{RTA_Gamma}.

The second-order equations of motion for $\Pi = -\frac{m_0^2}{3} \rho_0$, $V^\mu = \rho_0^\mu$ and 
$\pi^{\mu\nu} = \rho_0^{\mu\nu}$ follow after setting $r = 0$ in Eqs.~\eqref{D_rho_RTA}--\eqref{D_rho_munu_RTA}. 
Here, the positive-order moments vanish by the Landau-matching conditions and the choice of 
the Landau frame for the fluid velocity,  while the negative-order moments are only required up to first order, 
since they are always multiplied by terms of order $O(\Kn)$.

\subsection{Basis-free approach for the negative-order moments}
\label{sec:RTA:rhoneg}

A basis-free, first-order representation of the irreducible moments can be obtained directly from 
Eqs.~\eqref{D_rho_RTA}--\eqref{D_rho_munu_RTA}: 
\begin{align}
 \rho_r &\simeq \tau_R \alpha^{(0)}_r \theta\;, &
 \rho^\mu_r &\simeq \tau_R \alpha^{(1)}_r \nabla^\mu \alpha\;, &
 \rho^{\mu\nu}_r &\simeq 2\tau_R \alpha^{(2)}_r \sigma^{\mu\nu}\;,
 \label{bf_rho}
\end{align}
where all $O({\rm Re}^{-1}{\rm Kn})$ terms (including those of the type
$\tau_R \dot{\rho}^{\langle \mu_1 \cdots \mu_\ell \rangle}_r$) were neglected.
Expressing the thermodynamic forces $\theta$, $\nabla^\mu \alpha$, and $\sigma^{\mu\nu}$ in terms of the 
$r = 0$ moments leads to
\begin{gather}
	\rho_{r\neq 0} \simeq -\frac{3}{m_0^2} \mathcal{R}^{(0)}_{r0} \Pi\;, \nonumber\\
	\rho^\mu_{r \neq 0} \simeq \mathcal{R}^{(1)}_{r0} V^\mu\;,\quad
	\rho^{\mu\nu}_{r \neq 0} \simeq \mathcal{R}^{(2)}_{r0} \pi^{\mu\nu}\;, \label{bf_matching}
\end{gather}
where we have used Eq.~\eqref{R_def}.
When $r > 0$, employing Eq.~\eqref{main_RTA_coeffs} the relations~\eqref{bf_matching} are seen to be identical to the 
ones derived using the so-called IReD or order-of-magnitude approaches, shown in Eq.~\eqref{shifted_matching}.
Note that the relations~\eqref{bf_matching} are valid for any $r$, 
including $r < 0$, without having to calculate the negative-order moments through 
sums over moments of the chosen basis, such as those involved in computing $\gamma^{(\ell)}_{r0}$ and 
$\Gamma^{(\ell)}_{r0}$, hence leading to a direct basis-free approximation:
\begin{align}
 \rho_{-1} &\simeq -\frac{3}{m_0^2} \mathcal{R}^{(0)}_{-1,0} \Pi\;, & 
 \rho_{-2} &\simeq -\frac{3}{m_0^2} \mathcal{R}^{(0)}_{-2,0} \Pi\;, \label{bf_rho0neg}\\
 \rho^\mu_{-1} &\simeq \mathcal{R}^{(1)}_{-1,0} V^\mu\;, & 
 \rho^\mu_{-2} &\simeq \mathcal{R}^{(1)}_{-2,0} V^\mu\;, \label{bf_rho1neg}\\
 \rho^{\mu\nu}_{-1} &\simeq \mathcal{R}^{(2)}_{-1,0} \pi^{\mu\nu}\;, & 
 \rho^{\mu\nu}_{-2} &\simeq \mathcal{R}^{(2)}_{-2,0} \pi^{\mu\nu}\;.\label{bf_rho2neg}
\end{align}

\subsection{Shifted-basis approach for the negative-order moments}
\label{sec:sbanom}

We now consider the representation of the moments 
in the shifted-basis approach discussed in Sec.~\ref{sec:BTE:shifted}. 
For $ -s_\ell \leq r \leq N_\ell$, replacing the first-order transport coefficients in Eq.~\eqref{shifted_matching} 
by their RTA expression~\eqref{main_RTA_coeffs} reproduces Eq.~\eqref{bf_matching}. The moments
with $r < -s_\ell$ are still computed using Eq.~\eqref{shifted_rho_neg}.

When the mass $m_0 > 0$ and $s_\ell = 2$, the negative-order moments from 
Eqs.~\eqref{bf_rho0neg}--\eqref{bf_rho2neg} are identically reproduced.
In order to be able to apply the matching conditions $\rho_1 = \rho_2 = \rho_1^\mu = 0$, we have to make sure that
these moments are included in the basis. Thus, the truncation orders must satisfy 
\begin{equation}
 N_0 \ge 2, \quad N_1 \ge 1, \quad N_2 \ge 0\;.
\end{equation}
The smallest basis required to recover the RTA transport coefficients comprises 
$(N_0+s_0+1) \times 1 + (N_1+s_1+1) \times 3 + (N_2+s_2+1) \times 5 = 32$ 
moments. Accounting also for $n$, $e$, and $u^\mu$, there are a total of $37$ degrees of freedom, but
enforcing the matching conditions, this number is again brought down to 32.

In the case $m_0 = 0$, inspection of the equations of motion~\eqref{D_rho}--\eqref{D_rho_munu} 
for $r=0$ reveals that only the 
negative-order moments $\rho^\mu_{-1}$, $\rho^{\mu\nu}_{-1}$, and $\rho^{\mu\nu}_{-2}$ appear, which are perfectly 
compatible with the largest possible shift $s_\ell = \ell$. In this case, the smallest 
basis required to recover the RTA transport coefficients 
comprises $3 \times 1 + 3 \times 3 + 3 \times 5 = 27$ moments. 
The total number of degrees of freedom is then 32 
(including $n$, $e$, and $u^\mu$). This number is reduced by 5 due to the matching conditions and 
furthermore by 1, since the bulk viscous pressure vanishes for ultrarelativistic particles.

\subsection{Chapman-Enskog method}\label{sec:RTA:CE}

In this section, we employ the Chapman-Enskog method following Sec.\ 5.5 of 
Ref.~\cite{Cercignani_book} and establish the 
connection with the method of moments employed in this paper. 
The power-counting scheme is performed with respect 
to a parameter $\varepsilon \simeq \tau_R/L \sim O({\rm Kn})$ formally identified with the Knudsen number, such that
\begin{equation}
 \delta f_\bk \equiv f_\bk - f_{0\bk} =\varepsilon f^{(1)}_\bk + \varepsilon^2 f^{(2)}_\bk + \dots\;,\label{CE_df}
\end{equation}
while $f^{(0)}_\bk \equiv f_{0\bk}$ is the equilibrium distribution.

The collision term is assumed to be of order $O(\varepsilon^{-1})$, which is implemented in the RTA model by taking 
$\tau_R / \varepsilon$ to be of zeroth order with respect to $\varepsilon$.
The Boltzmann equation \eqref{BTE} in RTA, Eq.~(\ref{AW}), is then expanded as, 
cf.\ also Eq.~(28) of Ref.~\cite{Anderson:1974a}, 
\begin{equation}
 \sum_{i = 0}^\infty \varepsilon^i (k^\mu \partial_\mu f_\bk)^{(i)} 
 = -\frac{\varepsilon E_\bk}{\tau_R} \sum_{i = 0}^\infty \varepsilon^i f^{(i+1)}_\bk\;,
\label{BTE_CE_expansion}
\end{equation}
leading to an iterative procedure allowing $f^{(i+1)}_\bk$ to be obtained in terms of 
the lower-order terms $ f^{(j)}_\bk$ with $0 \le j \le i$.
The index $i$ of the expansion order takes into account the 
expansion of the comoving derivative, $D \equiv u^\mu \partial_\mu =  \sum_{j = 0}^{\infty} \varepsilon^j D_j$, such that 
the $i$th order contribution to the left-hand side of Eq.~\eqref{BTE_CE_expansion} reads:
\begin{equation}
 (k^\mu \partial_\mu f_\bk)^{(i)}
 = k^{\langle\mu\rangle} \nabla_\mu f_\bk^{(i)} + E_\bk \sum_{j = 0}^i D_j f_\bk^{(i- j)} \;. 
 \label{BTE_CE_rhs}
\end{equation}
The operator $D_j$ is introduced at the level of the thermodynamic variables $\alpha$, $\beta$, and $u^\mu$ via
\begin{gather}
 D\alpha = \sum_{j = 0}^\infty \varepsilon^j D_j \alpha\;, \quad
 D\beta = \sum_{j = 0}^\infty \varepsilon^j D_j\beta\;, \nonumber\\
 D u^\mu = \sum_{j = 0}^\infty \varepsilon^j D_j u^\mu\;,
\end{gather}
where the zeroth-order terms are
\begin{gather}
 D_0 \alpha = \frac{n \theta}{D_{20}}\left(h J_{20} - J_{30}\right)\;, \quad
 D_0 \beta = \frac{n \theta}{D_{20}}\left(h J_{10} - J_{20}\right)\;, \nonumber\\
 D_0 u^\mu = \frac{\nabla^\mu P}{e + P}\;,
 \label{CE_D0}
\end{gather}
while for $j > 0$, 
\begin{align}
 D_j \alpha &= \frac{J_{20}}{D_{20}} 
 \left[\Pi_{(j)} \theta - \pi_{(j)}^{\mu\nu} \sigma_{\mu\nu}\right] \nonumber\\
 & - \frac{J_{30}}{D_{20}} \left[\nabla_\mu V_{(j)}^\mu - 
 \sum_{i = 0}^{j-1} V^\mu_{(j-i)} D_i u_\mu\right]\;, \nonumber\\
 D_j \beta &= \frac{J_{10}}{D_{20}} 
 \left[\Pi_{(j)} \theta - \pi_{(j)}^{\mu\nu} \sigma_{\mu\nu}\right] \nonumber\\
 & - \frac{J_{20}}{D_{20}} \left[\nabla_\mu V_{(j)}^\mu - 
 \sum_{i = 0}^{j-1} V^\mu_{(j-i)} D_i u_\mu\right]\;, \nonumber\\
 D_j u^\mu &= \frac{\nabla^\mu \Pi_{(j)} - \Delta^\mu_\alpha \nabla_\beta \pi_{(j)}^{\alpha\beta}}{e + P} \nonumber\\
 &- \frac{1}{e + P}
 \sum_{i = 0}^{j-1} \left[\Pi_{(j-i)} D_i u^\mu - 
 \pi_{(j-i)}^{\mu\nu} D_i u_\nu\right]\;.\label{CE_Dj}
\end{align}

The first- and second-order corrections to $f_{0\bk}$ follow from Eq. (\ref{BTE_CE_expansion}), 
\begin{align}
 \varepsilon f^{(1)}_\bk &= -\frac{\tau_R}{E_\bk}
 \left[k^{\langle \mu\rangle} \nabla_\mu f_\bk^{(0)} + E_\bk D_0 f_\bk^{(0)}\right]\;,\label{CE_f1}\\
 \varepsilon^2 f^{(2)}_\bk &= -\varepsilon \frac{\tau_R}{E_\bk}
 \left[k^{\langle \mu\rangle} \nabla_\mu f_\bk^{(1)} + E_\bk D_0 f_\bk^{(1)}+ E_\bk D_1 f_\bk^{(0)}\right]\;.
 \label{CE_f2}
\end{align}
We now seek to reproduce the equation 
\begin{equation}
\delta  \dot{f}_\bk = - \dot{f}_{0\bk} - E_\bk^{-1} k_\nu \nabla^\nu f_{0\bk} - E_\bk^{-1} 
k_\nu \nabla^\nu \delta f_\bk + E_\bk^{-1} C[f]\;,
 \label{D_delta_f}
\end{equation}
which follows directly from the Boltzmann equation \eqref{BTE} (see Eq.~(34) in Ref.~\cite{Denicol:2012cn}). 
At leading order, the left-hand-side is $\delta \dot{f}_\bk \simeq \varepsilon D_0 f_\bk^{(1)}$, 
while the terms on the right-hand side can be approximated via:
\begin{align} 
 \dot{f}_{0\bk} &= D_0 f_\bk^{(0)} + \varepsilon D_1 f_\bk^{(0)}, &
 \frac{C[f]}{E_\bk} &= -\frac{\varepsilon}{\tau_R} 
 \left( f^{(1)}_\bk + \varepsilon f^{(2)}_{\bk} \right) \;.
 \label{CE_equivalence}
\end{align}
Employing Eqs.~\eqref{CE_f1}--\eqref{CE_f2}, it can be seen that Eq.~\eqref{D_delta_f} is recovered 
up to order $O(\varepsilon^1)$.
Since the moment equations \eqref{D_rho}--\eqref{D_rho_munu} are derived from Eq.~\eqref{D_delta_f}, 
the expressions in Eqs.~\eqref{CE_f1}--\eqref{CE_f2} will lead to the same equations, up to first order in 
$\varepsilon$. Upon multiplication with $\tau_R$, this is sufficient to derive the second-order equations of fluid dynamics. 
We note that the above conclusion was also established in Ref.~\cite{Mitra:2020gdk} for the tensor moments ($\ell = 2$).

The irreducible moments $\rho^{\mu_1 \cdots \mu_\ell}_r$ of $\delta f_\bk$ are written as
\begin{align}
 \rho^{\mu_1 \cdots \mu_\ell}_r &= \sum_{i=1}^\infty \varepsilon^i \rho^{\mu_1 \cdots \mu_\ell}_{r,(i)}\;, \label{CE_exp}\\
 \rho^{\mu_1 \cdots \mu_\ell}_{r,(i)} &= \int dK\, E_\bk^r k^{\langle \mu_1} \cdots k^{\mu_\ell \rangle} f^{(i)}_\bk\;.
 \label{CE_rhor}
\end{align}
The first-order contribution to the irreducible moments can be obtained using $f^{(1)}_\bk$ derived in Eq.~\eqref{CE_f1}, 
which can be written in explicit form by computing the comoving derivatives using Eqs.~\eqref{CE_D0}:
\begin{multline}
\varepsilon f^{(1)}_{\mathbf{k}}  
= \tau_R f_{0 \mathbf{k}} \bar{f}_{0\bk} \left\{ \frac{\beta \theta}{3 E_{\mathbf{k}}} \Delta^{\alpha\beta} 
k_\alpha k_\beta \right.\\
+ \frac{n\theta}{D_{20}} 
\left[J_{30} - h J_{20} + E_\bk (h J_{10} - J_{20})\right] \\
\left. +\left( \frac{1}{h} - \frac{1}{E_{\mathbf{k}} }\right) 
k^{\langle \mu \rangle}\nabla_{\mu}\alpha 
+\frac{\beta}{E_{\bk}} k^{\left\langle \mu \right. } 
k^{\left. \nu \right\rangle }\sigma_{\mu \nu }\right\}\;. \label{delta_f_CE}
\end{multline}
Plugging the above expressions into Eq.~\eqref{CE_rhor}, using the orthogonality relation (20) of
Ref.~\cite{Denicol:2012cn}, and focusing on the $\ell = 2$ case, we get
\begin{align}
\varepsilon \rho^{\mu \nu}_{r,(1)} &= \tau_R \beta \sigma^{\alpha \beta } 
\int dK f_{0 \mathbf{k}} \bar{f}_{0\bk} E_{\mathbf{k}}^{r-1} k^{\left\langle \mu \right. } 
k^{\left. \nu \right\rangle } k_{\left\langle \alpha \right. } 
k_{\left. \beta \right\rangle } \notag \\
&= 2 \tau_R \beta J_{r+3,2} \sigma^{\mu \nu } 
= 2  \tau_R \alpha^{(2)}_{r} \sigma^{\mu \nu }\;,
\label{CE_rho_munu}
\end{align}
where we employed $\beta J_{r+3,2} = \alpha^{(2)}_r$, which follows from Eqs.~(\ref{alpha2}) and (\ref{J_nq}).
Similarly,
\begin{multline}
	\varepsilon \rho^{\mu}_{r,(1)} = \tau_R \nabla^{\mu} \alpha 
	\int dK f_{0 \mathbf{k}} \bar{f}_{0\bk} E_{\mathbf{k}}^{r} \left( \frac{1}{h} - \frac{1}{E_{\mathbf{k}} }\right)  
	k^{\langle \nu \rangle } k_{\langle \mu \rangle } \\
	= \tau_R \left(J_{r+1,1}- \frac{J_{r+2,1}}{h} \right)\nabla^{\mu} \alpha 
	= \tau_R \alpha^{(1)}_{r} \nabla^{\mu} \alpha\;,
	\label{CE_rho_mu}
\end{multline}
where we used Eq.~\eqref{alpha1},
while with Eq.~\eqref{alpha0} the scalar moments reduce to
\begin{align}
\varepsilon \rho_{r,(1)} &= \tau_R \alpha^{(0)}_{r} \theta\;.
\label{CE_rho}
\end{align}
It can be seen that the first-order Chapman-Enskog results agree with those 
in Eq.~\eqref{bf_rho} obtained in the method of moments, hence the negative-order 
moments are also computed through Eqs.~\eqref{bf_rho0neg}--\eqref{bf_rho2neg}. 

In the RTA, the equivalence between the Chapman-Enskog method and the method of moments 
can be established also at 
second order by reproducing the equations of motion \eqref{D_rho}--\eqref{D_rho_munu}. For this purpose, 
the left-hand sides of the irreducible-moment equations can be expanded with respect to 
$\varepsilon$ using Eqs.~\eqref{CE_equivalence} and \eqref{CE_exp} as
\begin{multline}
 \dot{\rho}^{\langle \mu_1 \cdots \mu_\ell \rangle}_r - C^{\mu_1 \cdots \mu_\ell}_{r-1} =
 \frac{\varepsilon}{\tau_R} \rho_{r,(1)}^{\mu_1 \cdots \mu_\ell} \\
 + \varepsilon \left[D_0 \rho_{r,(1)}^{\mu_1 \cdots \mu_\ell} 
 + \frac{\varepsilon}{\tau_R} \rho_{r,(2)}^{\mu_1 \cdots \mu_\ell}\right] + O(\varepsilon^2)\;.
\end{multline}
The second-order contribution to the irreducible moments can be computed using Eqs.~\eqref{CE_f2} and
\eqref{CE_rhor},
\begin{multline}
 \varepsilon^2 \rho^{\mu_1 \cdots \mu_\ell}_{r,(2)} = 
 -\varepsilon \tau_R \Delta^{\mu_1 \cdots\mu_\ell}_{\nu_1 \cdots \nu_\ell}
 \int dK \, E_\bk^r k^{\langle\nu_1} \cdots k^{\nu_\ell \rangle}\\\times
 \left[D_0 f_\bk^{(1)} + D_1 f_\bk^{(0)} + \frac{k^{\langle \mu\rangle}}{E_\bk} \nabla_\mu f_\bk^{(1)}\right]\;.
\end{multline}
Taking the comoving derivative $D_0$ outside the integral provides 
$D_0 \rho^{\langle \mu_1 \cdots \mu_\ell \rangle}_{r,(1)}$, such that 
\begin{multline}
 D_0 \rho^{\langle \mu_1 \cdots \mu_\ell\rangle}_{r,(1)} + \frac{\varepsilon}{\tau_R} \rho^{\mu_1 \cdots \mu_\ell}_{r,(2)} \\
 = \Delta^{\mu_1 \cdots\mu_\ell}_{\nu_1 \cdots \nu_\ell}
 \int dK \, \left[ D_0\left(E_\bk^r k^{\langle\nu_1} \cdots k^{\nu_\ell\rangle}\right) \right] f_\bk^{(1)}\\
 - \int dK \, E_\bk^r k^{\langle\mu_1} \cdots k^{\mu_\ell \rangle}\left[D_1 f_\bk^{(0)} 
 + \frac{k^{\langle \mu\rangle}}{E_\bk} \nabla_\mu f^{(1)}_\bk\right]\;.
\end{multline}
The right-hand side of the above expression together with the Navier-Stokes contribution from 
$\rho^{\mu_1 \cdots \mu_\ell}_{r,(1)}$ generate all of the terms appearing on the right-hand sides of 
Eqs.~\eqref{D_rho}--\eqref{D_rho_munu}. 

Discrepancies between the results obtained using the Chapman-Enskog method and the method of moments 
were reported in the literature at the level of the second-order transport coefficients. These discrepancies are in 
fact due to the omission of certain second-order terms, as we point out in detail in Appendix~\ref{app:CE}.

\subsection{Transport coefficients in the 14-moment approximation}
\label{sec:RTA:tcoeffs}

Here we recall the general form of the second-order transport equations for 
$\Pi$, $V^\mu$, and $\pi^{\mu\nu}$ from Ref.~\cite{Denicol:2012cn},
\begin{align}
	\tau_\Pi \dot{\Pi} + \Pi &=  -\zeta \theta + \mathcal{J} + \mathcal{K} + \mathcal{R}\;,\label{Pidot}\\
	\tau_V \dot{V}^{\langle \mu \rangle} + V^\mu &= \kappa \nabla^\mu \alpha + 
	\mathcal{J}^\mu + \mathcal{K}^\mu + \mathcal{R}^\mu\;, \label{Vdot}\\
	\tau_\pi \dot{\pi}^{\langle \mu \nu \rangle} + \pi^{\mu\nu} &= 
	2 \eta \sigma^{\mu\nu} + \mathcal{J}^{\mu\nu} + \mathcal{K}^{\mu\nu} + 
	\mathcal{R}^{\mu\nu}\;,
	\label{pidot}
\end{align}
where $\tau_\Pi$, $\tau_V$, and $\tau_\pi$ are the relaxation 
times, $\zeta = \zeta_0$, $\kappa = \kappa_0$, and $\eta=\eta_0$ 
are the first-order transport coefficients, while
$\mathcal{J}, \mathcal{J}^{\mu}$, and $\mathcal{J}^{\mu\nu}$ 
collect terms of order $O({\rm Re}^{-1} \rm{Kn})$:
\begin{align}
	\mathcal{J} &= -\ell_{\Pi V} \nabla_\mu V^\mu - 
	\tau_{\Pi V} V_\mu \dot{u}^\mu - \delta_{\Pi\Pi} \Pi \theta \nonumber\\
	& - \lambda_{\Pi V} V_\mu \nabla^\mu \alpha + 
	\lambda_{\Pi \pi} \pi^{\mu\nu} \sigma_{\mu\nu}\;,\label{J}\\
	\mathcal{J}^\mu &= -\tau_V V_\nu \omega^{\nu\mu} - \delta_{VV} V^\mu \theta 
	- \ell_{V\Pi} \nabla^\mu \Pi \nonumber\\
	& + \ell_{V\pi} \Delta^{\mu\nu} \nabla_\lambda \pi^\lambda{}_\nu + \tau_{V\Pi} \Pi \dot{u}^\mu - 
	\tau_{V\pi} \pi^{\mu\nu} \dot{u}_\nu \nonumber\\
	& -\lambda_{VV} V_\nu \sigma^{\mu\nu} + 
	\lambda_{V \Pi} \Pi \nabla^\mu \alpha - \lambda_{V\pi} \pi^{\mu\nu} \nabla_\nu \alpha\;,\label{J_mu}\\
	\mathcal{J}^{\mu\nu} &= 2\tau_\pi \pi^{\langle\mu}_\lambda \omega^{\nu\rangle \lambda} - 
	\delta_{\pi\pi} \pi^{\mu\nu} \theta - 
	\tau_{\pi\pi} \pi^{\lambda\langle \mu}\sigma^{\nu\rangle}_\lambda + 
	\lambda_{\pi \Pi} \Pi \sigma^{\mu\nu} \nonumber\\
	& - \tau_{\pi V} V^{\langle \mu} \dot{u}^{\nu \rangle} + \ell_{\pi V} \nabla^{\langle \mu} V^{\nu \rangle} 
	+ \lambda_{\pi V} V^{\langle \mu} \nabla^{\nu \rangle} \alpha\;.
	\label{J_munu}
\end{align}
The tensors $\mathcal{K}$, $\mathcal{K}^\mu$, and 
$\mathcal{K}^{\mu\nu}$ contain ${\rm Kn}^2$ contributions,
which will play no role in the following.
The tensors $\mathcal{R}$, $\mathcal{R}^\mu$, and $\mathcal{R}^{\mu\nu}$ 
contain terms of order ${\rm Re}^{-2}$ originating from quadratic terms 
in the collision integral, which are absent in RTA.

We are now ready to determine the transport coefficients. For the sake of definiteness,
we work within the basis-free approach and note that similar results 
are obtained when using the shifted-basis approach. 
The results obtained using the DNMR and corrected DNMR approaches can be 
obtained by replacing
\begin{align}
\mathcal{R}^{(\ell)}_{-r,0} &\rightarrow \gamma^{(\ell)}_{r0}\;, \qquad 
\mathcal{R}^{(\ell)}_{-r,0} \rightarrow \Gamma^{(\ell)}_{r0}\;.
\label{comparison}
\end{align}

While in the RTA, the relaxation times satisfy
\begin{equation}
	\tau_\Pi = \tau_V = \tau_\pi  = \tau_R\;,
\end{equation}
we will use $\tau_\Pi$, $\tau_V$, and $\tau_\pi$ explicitly for the sake of clarity.
The transport coefficients appearing in the equation for the bulk viscous pressure are:
\begin{align}
	\zeta &= \tau_\Pi\frac{m_0^2}{3} \alpha^{(0)}_{r}\;, \label{bf_zeta} \\
	\delta_{\Pi\Pi} &= \tau_\Pi \left[\frac{2}{3} - \frac{m_0^2}{3} \frac{G_{20}}{D_{20}} 
	+ \frac{m_0^2}{3} \mathcal{R}^{(0)}_{-2,0}\right]\;, \\	
	\ell_{\Pi V} &= \tau_\Pi\frac{m_0^2}{3}\left[\frac{G_{30}}{D_{20}} - \mathcal{R}^{(1)}_{-1,0} \right] \;, \\
	\tau_{\Pi V} &= - \tau_\Pi\frac{m_0^2}{3} \left[\frac{G_{30}}{D_{20}} 
	- \frac{\partial \mathcal{R}^{(1)}_{-1,0} }{\partial \ln \beta} \right]\;, \\ 
	\lambda_{\Pi V} &= -\tau_\Pi\frac{m_0^2}{3} 
	\left[\frac{\partial \mathcal{R}^{(1)}_{-1,0}}{\partial \alpha} + 
	\frac{1}{h} \frac{\partial \mathcal{R}^{(1)}_{-1,0}}{\partial \beta}\right]\;, \\
	\lambda_{\Pi \pi} &=  -\tau_\Pi\frac{m_0^2}{3} \left[\frac{G_{20}}{D_{20}} 
	-\mathcal{R}^{(2)}_{-2,0}\right]\;. 
\end{align}
The transport coefficients for the diffusion equation are
\begin{align}
	\kappa &= \tau_V \alpha^{(1)}_0\;, \quad \label{bf_kappa}
	\delta_{VV} = \tau_V \left[1 + \frac{m_0^2}{3} \mathcal{R}^{(1)}_{-2,0} \right]\;, \\
	\ell_{V \Pi} &= \frac{\tau_V}{h}\! \left[1 -h \mathcal{R}^{(0)}_{-1,0}\right]\;, \
	\ell_{V \pi} = \frac{\tau_V}{h} \!\left[ 1- h\mathcal{R}^{(2)}_{-1,0}\right]\;, \\
	\tau_{V \Pi} &= \frac{\tau_V}{h}\! \left[1 - 
	h \frac{\partial \mathcal{R}^{(0)}_{-1,0}}{\partial \ln \beta}\right]\;, \ 
	\tau_{V \pi} = \frac{\tau_V}{h} \! \left[1
	- h\frac{\partial \mathcal{R}^{(2)}_{-1,0}}{\partial \ln \beta}\right]\;, \\
	\lambda_{VV} &= \tau_V \left[\frac{3}{5} + \frac{2m_0^2}{5} \mathcal{R}^{(1)}_{-2,0} \right]\;, \\
	\lambda_{V \Pi} &= \tau_V \left[\frac{\partial \mathcal{R}^{(0)}_{-1,0}}{\partial \alpha} 
	+ \frac{1}{h} \frac{\partial \mathcal{R}^{(0)}_{-1,0}}{\partial \beta} \right]\;, \\
	\lambda_{V \pi} &= \tau_V \left[\frac{\partial \mathcal{R}^{(2)}_{-1,0}}{\partial \alpha} 
	+ \frac{1}{h} \frac{\partial \mathcal{R}^{(2)}_{-1,0}}{\partial \beta} \right]\;. \label{bf_lambda_V_pi}
\end{align}
Finally, the transport coefficients appearing in the equation for the shear-stress tensor are
\begin{align}
	\eta &= \tau_\pi \alpha^{(2)}_0\;, \label{bf_eta} \quad
	\delta_{\pi\pi} = \tau_\pi \left[ \frac{4}{3} + \frac{m_0^2}{3} \mathcal{R}^{(2)}_{-2,0} \right]\;,\\
	\tau_{\pi\pi} &= \tau_\pi \left[\frac{10}{7} + \frac{4m_0^2}{7} \mathcal{R}^{(2)}_{-2,0} \right]\;,\\
	\lambda_{\pi \Pi} &= \tau_\pi\left[\frac{6}{5} + \frac{2m_0^2}{5} \mathcal{R}^{(0)}_{-2,0} \right]\;,\\
	\tau_{\pi V} &= -\tau_\pi \frac{2m_0^2}{5} 
	\frac{\partial \mathcal{R}^{(1)}_{-1,0}}{\partial \ln \beta}\;, \quad
	\ell_{\pi V} =  -\tau_\pi\frac{2m_0^2 }{5} \mathcal{R}^{(1)}_{-1,0}\;, \\
	\lambda_{\pi V} &=  -\tau_\pi\frac{2m_0^2 }{5} 
	\left[\frac{\partial \mathcal{R}^{(1)}_{-1,0}}{\partial \alpha} + 
	\frac{1}{h} \frac{\partial \mathcal{R}^{(1)}_{-1,0}}{\partial \beta}\right]\;.
	\label{bf_lambda_pi_V}
\end{align}

One also observes that when $m_0 > 0$, all coefficients except the first-order ones, 
$\zeta$, $\kappa$, and $\eta$, involve the functions $\mathcal{R}^{(\ell)}_{-r,0}$. 
These are related to the representation of the negative-order moments, as indicated in Eq.~\eqref{comparison}.

\subsection{Magnetohydrodynamics transport coefficients}
\label{sec:RTA:MHD}

Here we also consider the transport coefficients arising from the Boltzmann-Vlasov equation using
the method of moments as derived in Refs.~\cite{Denicol.2018,Denicol:2019iyh}, 
leading to the equations of non-resistive and resistive magnetohydrodynamics. 
Without repeating the details presented there, we summarize the additional 
$\mathcal{J}_{em}^{\mu_1 \cdots \mu_\ell}$ terms that appear on the right-hand sides of 
Eqs.~\eqref{J}--\eqref{J_munu} due to the coupling of the electric charge $\mathfrak{q}$ 
to the electromagnetic field
\begin{align}
 \mathcal{J}_{em} &= -\mathfrak{q} \delta_{\Pi V E} V^\nu E_\nu\;,\label{MHD_J}\\
 \mathcal{J}_{em}^\mu &= \mathfrak{q}\left(\delta_{VE} E^\mu + \delta_{V \Pi E} \Pi E^\mu
 + \delta_{V\pi E} \pi^{\mu\nu} E_\nu \right)  \nonumber\\
 &- \mathfrak{q} \delta_{VB} B b^{\mu\nu} V_\nu \;,\label{MHD_J_mu}\\
 \mathcal{J}_{em}^{\mu\nu} &= -\mathfrak{q} \left(\delta_{\pi B} B b^{\alpha\beta} 
 \Delta^{\mu\nu}_{\alpha\kappa} \pi^\kappa{}_\beta + \delta_{\pi V E} E^{\langle \mu} V^{\nu \rangle}\right)\;.
\end{align}
These are obtained from Eqs.~(24)--(26) of Ref.~\cite{Denicol:2019iyh} by employing the Landau frame, i.e., 
$W^\mu \equiv \rho_1^\mu = 0$. In the above the electric 
and magnetic fields $E^\mu$ and $B^\mu$ are defined through the Faraday tensor $F^{\mu\nu}$ and
the fluid four-velocity $u^\mu$ via 
\begin{equation}
 E^\mu = F^{\mu\nu}u_\nu \;, \quad 
 B^\mu = \frac{1}{2} \epsilon^{\mu\nu \alpha\beta} F_{\alpha\beta} u_\nu\;,
\end{equation}
while $b^{\mu\nu} = - \epsilon^{\mu\nu\alpha\beta} u_\alpha b_\beta$,  $b^\mu = B^\mu /B $, and 
$B = \sqrt{-B^\mu B_\mu}$ is the magnitude of the magnetic field.

The corresponding transport coefficients proportional to the electric and magnetic fields are obtained by replacing 
$(\tau^{(0)}_{00}, \tau^{(1)}_{00}, \tau^{(2)}_{00}) \rightarrow (\tau_\Pi, \tau_V,\tau_\pi)$ 
and $\gamma^{(\ell)}_r \rightarrow \mathcal{R}^{(\ell)}_{-r,0}$. These are
\begin{align}
	\delta_{VE} &= \tau_V\left( -\frac{n}{h} + \beta J_{11}\right)\; , \label{bf_delta_V_E} \\
	\delta_{\Pi VE}&= -\tau_\Pi\frac{m_{0}^{2}}{3}
	\left[\frac{G_{20}}{D_{20}} -\mathcal{R}^{(1)}_{-2,0}
	+\frac{1}{h} \frac{\partial \mathcal{R}^{(1)}_{-1,0}}{\partial \ln\beta} \right]\;, \\
	\delta_{V\Pi E} &= -\tau_V \left[ \frac{2}{m_{0}^{2}}+\mathcal{R}^{(1)}_{-2,0}
	- \frac{1}{h} \frac{\partial \mathcal{R}^{(0)}_{-1,0}}{\partial \ln\beta} \right]\;, \\
	\delta_{V\pi E} &=\tau_V \left[\mathcal{R}^{(2)}_{-2,0} - \frac{1}{h} 
	\frac{\partial \mathcal{R}^{(2)}_{-1,0}}{\partial \ln\beta} \right]\;,\\
	\delta_{\pi VE} &= \tau_\pi\left[ \frac{8}{5} + \frac{2m_{0}^{2}}{5} \mathcal{R}^{(1)}_{-2,0} 
	- \frac{2m_{0}^{2}}{5h}
	\frac{\partial \mathcal{R}^{(1)}_{-1,0}}{\partial \ln\beta} \right]\;, \label{bf_delta_pi_V_E}
\end{align}
and
\begin{align}
	\delta_{VB} &=\tau_V \left[ -\frac{1}{h} + \mathcal{R}^{(1)}_{-1,0}\right]\;, &
	\delta_{\pi B} &= 2\tau_\pi \mathcal{R}^{(2)}_{-1,0}\;. \label{bf_delta_pi_B}
\end{align}

\section{Results for the ideal ultrarelativistic Boltzmann gas}\label{sec:UR}

In this section, we analyze the classical, ultrarelativistic limit of the transport coefficients listed in 
Eqs.~\eqref{bf_zeta}--\eqref{bf_delta_pi_B}. In this limit, the bulk viscous pressure $\Pi$ vanishes
and all related transport coefficients do not need to be considered.
We begin this section with an explicit computation of the thermodynamic functions 
and the polynomial basis focusing on the specific case $s_0 = s_1 = s_2 = 0$.
We then compute the functions $\mathcal{F}^{(\ell)}_{rn}$, as well as the coefficients 
$\gamma^{(\ell)}_{r0} = \mathcal{F}^{(\ell)}_{r0}$, cf.\ Eq.~\eqref{RTA_gamma}, and $\Gamma^{(\ell)}_{r0}$,
cf.\ Eq.~\eqref{RTA_Gamma} with $s_\ell =0$ (in which case $\Gamma^{(\ell)}_{r0} = \widetilde{\Gamma}^{(\ell)}_{r0}$).
Finally, we report the transport coefficients.

\subsection{Thermodynamic functions}\label{sec:UR:thermo}

The equilibrium distribution of an ideal Boltzmann gas is obtained by setting $a = 0$ in Eq.~\eqref{f_0k} and
corresponds to the Maxwell-J\"uttner distribution:
\begin{align}
 \feq =& e^{\alpha -\beta E_\bk}\;.
 \label{UR_feq}
\end{align}
Since $\partial \feq/\partial \alpha  = \feq$,
$J_{nq} = I_{nq}$ by virtue of Eq.~\eqref{J_nq}.
The $I_{nq}$ integrals can be expressed in terms of the pressure 
$P = g e^\alpha / \pi^2 \beta^4$ as
\begin{equation}
 I_{nq} = \frac{P \beta^{2 - n}}{2(2q+1)!!} (n+1)!\;.
 \label{UR_Inq}
\end{equation}
Using this result in Eqs.~(\ref{alpha1}) and (\ref{alpha2}) gives
\begin{align}
 \alpha^{(1)}_r &= \frac{P(r+2)!(1-r)}{24 \beta^{r-1}}\;, &
 \alpha^{(2)}_r &= \frac{P}{30\beta^r} (r + 4)!\;,
 \label{UR_alpha_12}
\end{align}
allowing us to express the ratios $\mathcal{R}^{(\ell)}_{r0}$ from Eq.~\eqref{R_def} as
\begin{align}
 \mathcal{R}^{(1)}_{r0} &= \frac{(r+2)!(1-r)}{2\beta^r}\;, &
 \mathcal{R}^{(2)}_{r0} &= \frac{(r+4)!}{24\beta^r}\;.
 \label{UR_Rr}
\end{align}
Therefore, when $r = -1,-2$ the above results reduce to
\begin{align}
 \mathcal{R}^{(1)}_{-1,0} &= \beta\;,&
 \mathcal{R}^{(1)}_{-2,0} &= \frac{3\beta^2}{2}\;,
 \label{UR_R1neg}\\
 \mathcal{R}^{(2)}_{-1,0} &= \frac{\beta}{4}\;, &
 \mathcal{R}^{(2)}_{-2,0} &= \frac{\beta^2}{12}\;.
 \label{UR_R2neg}
\end{align}

\subsection{Polynomial basis} \label{sec:UR:polys}

We now construct the polynomials $P_{\bk m}^{(\ell)}$ and $\mathcal{H}^{(\ell)}_{\bk n}$ 
for the case $s_0 = s_1 = s_2 = 0$ considered in Ref.~\cite{Denicol:2012cn}. By the convention 
of Sec.~\ref{sec:BTE:delta_f_expansion} the overhead tildes are omitted. 
 Substituting Eq.~\eqref{UR_Inq} for $I_{nq}$ into Eq.~\eqref{eq:omega_def}, we find
\begin{equation}
\omega^{(\ell)} = \frac{2 \beta^{2\ell - 2} E_\bk^{2\ell}}{P(2\ell + 1)!} \feq\;,
\label{DNMR_UR_omega}
\end{equation}
where $(-\Delta^{\alpha\beta} k_\alpha k_\beta)^\ell = E_\bk^{2\ell}$ in the ultrarelativistic limit, $m_0 = 0$.
Plugging this into the orthogonality relation \eqref{eq:P_ortho} with $E_\bk = x/\beta$ gives
\begin{equation}
 \frac{1}{(2\ell + 1)!} 
 \int_0^\infty dx\, e^{-x} 
 x^{2\ell + 1} P^{(\ell)}_{\bk m}\left(\frac{x}{\beta}\right) 
 P^{(\ell)}_{\bk n}\left(\frac{x}{\beta}\right) 
 = \delta_{mn}\;.\label{UR_P_ortho}
\end{equation}
The above relation is similar to the orthogonality relation obeyed by the generalized Laguerre polynomials,
\begin{multline}
 \int_0^\infty dx\, e^{-x} x^{2\ell + 1} 
 L_m^{(2\ell+1)}(x) L_n^{(2\ell+1)}(x) \\
 = \frac{(n+2\ell+1)!}{n!} \delta_{mn}\;.
 \label{Laguerre_ortho}
\end{multline}
Based on this analogy, the polynomials $P^{(\ell)}_{\bk m}(E_\bk)$ can be expressed in 
terms of the generalized Laguerre polynomials $L^{(2\ell+1)}_m(\beta E_\bk)$ as
\begin{equation}
 P^{(\ell)}_{\bk m}(E_\bk) = \sqrt{\frac{m!(2\ell+1)!}{(m+2\ell+1)!}}
 L^{(2\ell+1)}_m(\beta E_\bk)\;.
 \label{UR_P}
\end{equation}
Given the explicit representation 
\begin{equation}
 L^{(2\ell + 1)}_m(x) = \sum_{n = 0}^m
 \frac{(-x)^n(m + 2\ell + 1)!}{n! (m-n)! (n + 2\ell + 1)!}\;,
 \label{Laguerre_expl}
\end{equation}
the expansion coefficients $a_{mn}^{(\ell)}$ appearing in the representation of 
$P^{(\ell)}_{\bk m}(E_\bk)$ from Eq.~\eqref{eq:P_k} are identified as
\begin{equation}
 a_{mn}^{(\ell)} = (-\beta)^n
 \frac{\sqrt{m!(2\ell+1)!(m+2\ell+1)!}}
 {n!(m-n)!(n+2\ell+1)!}\;.
 \label{UR_a}
\end{equation}

\subsection{DNMR coefficients $\gamma^{(\ell)}_{r0}$}\label{sec:UR:gamma}

In this subsection we obtain a closed form for the coefficients 
$\gamma^{(\ell)}_{r0} = \mathcal{F}^{(\ell)}_{r0}$. Starting from Eq.~\eqref{F_rn}, we set
$s_\ell = 0$ and $J_{nq} = I_{nq}$, with $I_{nq}$ from Eq.~\eqref{UR_Inq}, and
use Eq.~\eqref{UR_a} for the coefficients $a^{(\ell)}_{mn}$ and $a^{(\ell)}_{mq}$, which ultimately leads to
\begin{equation}
 \mathcal{F}^{(\ell)}_{rn} = \frac{(-1)^n \beta^{r+n}}{(n+2\ell+1)!} 
 \sum_{m = n}^{N_\ell} \frac{(m + 2\ell + 1)!}{n!(m-n)!} 
 \mathcal{S}_m\;,
 \label{F_rn_aux}
\end{equation}
where we introduced 
\begin{align}
 \mathcal{S}_m \equiv& \sum_{q = 0}^m (-1)^q \binom{m}{q} 
 \frac{(q + 2\ell + 1 - r)!}{(q + 2\ell + 1)!}\;.
 \label{Smab}
\end{align}
In order to find $\mathcal{S}_m$, we recall the definition of the Gauss hypergeometric function \cite{olver2010nist},
\begin{equation}
 {}_2F_1(a,b;c;z) = \sum_{q = 0}^\infty \frac{(a)_q (b)_q}{(c)_q q!} z^q\;,\label{2F1}
\end{equation}
where $(a)_q = \Gamma(a + q) / \Gamma(a)$ is the Pochhammer symbol. Using the property
\begin{equation}
 (-m)_q = (-1)^q\frac{m!}{(m-q)!}\;,
\end{equation}
valid for $m,q \ge 0$, we get
\begin{equation}
 \mathcal{S}_m = \frac{(2\ell + 1 - r)!}{(2\ell+1)!} {}_2F_1(-m,2\ell + 2-r;2\ell + 2; 1)\;.
\end{equation}
Note that the summation in Eq.~\eqref{2F1} is truncated at $q = m$ since $m!/(m-q)!$ 
vanishes when $q > m$. Using now the identity \cite{olver2010nist} 
\begin{equation}
 {}_2F_1(-m,b;c;1) = \frac{(c-b)_m}{(c)_m}\;,
\end{equation}
we arrive at
\begin{equation}
 \mathcal{S}_m = \frac{ (2\ell + 1 -r)! (r-1+m)!}{(2\ell+1 + m)!(r - 1)!}\;.
 \label{Smab_sol}
\end{equation}

Substituting Eq.~\eqref{Smab_sol} into Eq.~\eqref{F_rn_aux} leads to
\begin{equation}
 \mathcal{F}^{(\ell)}_{rn} = \frac{\beta^{r+n}}{r+n} \frac{(-1)^n (2\ell + 1 - r)! (N_\ell + r)!}{n!(r-1)!(2\ell + 1 + n)! (N_\ell - n)!}\;,
 \label{UR_F_rn}
\end{equation}
which is valid when $r \le 2\ell + 1$. When $r > 2\ell + 1$, the integral in Eq.~\eqref{F_rn} becomes infrared divergent in 
the massless limit, due to the negative power of $E_\bk$. However, the only moments $\rho_{-r}^{\mu_1 \cdots \mu_\ell}$
which enter the equations of motion are those with $r \leq 2$, see Eq.~\eqref{needed_rhoneg_m}.
In the massless limit, the scalar $(\ell =0)$ moments
$\rho_{-2}, \rho_{-1}$, and $\rho_0$ are not considered, so we do not need to discuss this case 
any further. On the other hand, for the vector $(\ell = 1)$ and tensor $(\ell = 2)$ moments this problem does not arise, 
since there $r \leq 2 < 2\ell +1$.

The validity of Eq.~\eqref{UR_F_rn} can also be extended to $r \le 0$, by replacing $(r+n)(r-1)!$ in the denominator by 
$(r+n)\Gamma(r)$. Since $\Gamma(r)$ has simple poles when $r = 0, -1, -2, \dots$ is a non-positive integer, 
$\mathcal{F}^{(\ell)}_{rn}$ vanishes whenever $r \le 0$ and $r + n \neq 0$. The value of $\mathcal{F}^{(\ell)}_{-n,n}$ 
can be obtained by taking the limit $r \rightarrow -n$ using 
\begin{equation}
 \lim_{r \rightarrow -n} (r+n) \Gamma(r) = \frac{(-1)^n}{n!}\;.
\end{equation}
Substituting the above into Eq.~\eqref{UR_F_rn}  gives
\begin{equation}
 \mathcal{F}_{-r,n}^{(\ell)}= \delta_{rn} \;,
\end{equation}
for $N_\ell \ge r \ge 0$, which is the expected result, see discussion after Eq.~\eqref{useful}.

For $\ell = 1,2$ and $r = 1,2$, the functions $\gamma^{(\ell)}_{r0} = \mathcal{F}^{(\ell)}_{r0}$ are obtained as
\begin{align}
 \gamma^{(1)}_{10} =& \frac{\beta (1 + N_1)}{3}\;, &
 \gamma^{(1)}_{20} =& \frac{\beta^2 (1 + N_1)(2 + N_1)}{12}\;,
 \label{UR_gamma1}\\
 \gamma^{(2)}_{10} =& \frac{\beta (1 + N_2)}{5}\;, & 
 \gamma^{(2)}_{20} =& \frac{\beta^2 (1 + N_2)(2 + N_2)}{40}\;.
 \label{UR_gamma2}
\end{align}
The above expressions diverge for $N_\ell \rightarrow \infty $.
However, in the following we will show that the corrected DNMR coefficients $\Gamma^{(\ell)}_{r0}$ do not diverge
and, at least in RTA, actually agree with $\mathcal{R}^{(\ell)}_{-r,0}$ listed in Eqs.~(\ref{UR_R1neg}) and (\ref{UR_R2neg}).

\subsection{Corrected DNMR coefficients $\Gamma^{(\ell)}_{r0}$}\label{sec:UR:Gamma}

We now compute the corrected coefficients $\Gamma^{(\ell)}_{r0}$ in the RTA from Eq.~\eqref{RTA_Gamma}. 
Employing the expressions \eqref{UR_Rr} and \eqref{UR_F_rn} for $\mathcal{R}^{(\ell)}_{n0}$ and 
$\mathcal{F}^{(\ell)}_{rn}$, respectively, gives
\begin{align}
 \Gamma^{(1)}_{r0} =& \frac{\beta^r (3 - r)! (N_1 + r)!}{2(r-1)! N_1!}
 \sum_{n = 0}^{N_1} \binom{N_1}{n} \frac{(-1)^n (1 - n)}{(n+3)(n+r)}\;,\label{UR_Gamma1_aux}\\
 \Gamma^{(2)}_{r0} =& \frac{\beta^r (5 - r)! (N_2 + r)!}{24(r-1)! N_2!}
 \sum_{n = 0}^{N_2} \binom{N_2}{n} \frac{(-1)^n}{(n+5)(n+r)}\;.
 \label{UR_Gamma_2_aux}
\end{align}
Defining the functions
\begin{align}
 S_N(x;a) & \equiv \sum_{n = 0}^N \binom{N}{n} \frac{(-x)^n}{n+a+1} \;,\\
 S_N(x;a,b) &\equiv \sum_{n = 0}^N \binom{N}{n} \frac{(-x)^n}{(n+a+1)(n+b+1)}\;,
 \label{UR_SN_def}
\end{align}
we can express the coefficients (\ref{UR_Gamma1_aux}), (\ref{UR_Gamma_2_aux}) as
\begin{align}
 \Gamma^{(1)}_{r0} &= \frac{\beta^r (3-r)! (r + N_1)!}{2(r-1)! N_1!} \nonumber \\
  & \times [4S_{N_1}(1;2,r-1) - S_{N_1}(1;r-1)]\;, \\ 
 \Gamma^{(2)}_{r0} &= \frac{\beta^r (5-r)! (r + N_2)!}{24(r-1)! N_2!} 
 S_{N_2}(1;4,r-1)\;.
 \end{align}
 
The functions $S_N(x;a)$ and $S_N(x;a,b)$ have an integral representation
 \begin{align}
 S_N(x;a) &  = \frac{1}{x^{a+1}} \int_0^x dt\, t^a S_N(t)\;,\\
 S_N(x;a,b) & = \frac{1}{x^{b+1}} \int_0^x dt\, t^b S_N(t;a)\;,
 \label{UR_SN_int}
\end{align}
where 
\begin{equation}
 S_N(x) \equiv \sum_{n = 0}^N \binom{N}{n} (-x)^n = (1 - x)^N \;,
\end{equation}
by the binomial theorem.
Using the definition of the incomplete Beta function,
\begin{equation}
 B_x(a,b) = \int_0^x dt\, t^{a-1} (1 - t)^{b-1}\;,
 \label{Beta}
\end{equation}
one immediately concludes that
\begin{equation}
 S_N(x;a) = \frac{1}{x^{a+1}} B_x(a+1,N+1)\;.
 \label{UR_SN}
\end{equation}
Setting $x = 1$ in the above expression, $B_x(a,b)$ becomes the complete Beta function $B(a,b)$ \cite{olver2010nist},
\begin{equation}
 B_1(a,b) \equiv B(a,b) = \frac{\Gamma(a)\Gamma(b)}{\Gamma(a+b)}\;,
\end{equation}
such that
\begin{equation}
 S_N(1;a) = \sum_{n = 0}^N \binom{N}{n} \frac{(-1)^n}{n + a + 1} =
 \frac{a! N!}{(N + a + 1)!}\;.
\end{equation}
In the case of $S_N(x;a,b)$, we can consider directly the case $x = 1$ to find
\begin{multline}
 S_N(1;a,b) = \int_0^1 dx \, x^{b - a - 1} B_x(a+1,N+1)\\
 = \frac{1}{a-b}[B(b+1, N+1) - B(a+1, N+1)]\;.
\end{multline}
With the above, we arrive at
\begin{align}
 \Gamma^{(1)}_{r0} &  = \frac{\beta^r(2-r)!(r+1)}{2} 
 \left[1 - \frac{8 r (r + N_1)!}{(r+1)!(3 + N_1)!}\right]\;,
 \label{UR_Gamma1} \\
 \Gamma^{(2)}_{r0} & = \frac{\beta^r(4-r)!}{24} 
 \left[1 - \frac{24 (r + N_2)!}{(r-1)!(5 + N_2)!}\right]\;.
 \label{UR_Gamma2}
\end{align}
In the limit $N_1, N_2 \rightarrow \infty$, 
$\Gamma^{(1)}_{r0}$ and $\Gamma^{(2)}_{r0}$ reduce to 
$\mathcal{R}^{(1)}_{-r,0}$ and $\mathcal{R}^{(2)}_{-r,0}$ given in Eqs.~\eqref{UR_Rr}:
\begin{equation}
 \lim_{N_1 \rightarrow \infty} \Gamma^{(1)}_{r0} = \mathcal{R}^{(1)}_{-r,0}\;,\qquad 
 \lim_{N_2 \rightarrow \infty} \Gamma^{(2)}_{r0} = \mathcal{R}^{(2)}_{-r,0}\;.
\end{equation}

Setting now $r = 1$ and $2$ leads to 
\begin{align}
 \Gamma^{(1)}_{10} =& \beta \left[1 - \frac{4}{(2 + N_1)(3 + N_1)}\right]\;,\label{UR_Gamma11}\\
 \Gamma^{(1)}_{20} =& \frac{3\beta^2}{2} \left[1 - \frac{8}{3(3 + N_1)}\right]\;,\label{UR_Gamma12}\\
 \Gamma^{(2)}_{10} =& \frac{\beta}{4} \left[1 - \frac{24(1 + N_2)!}{(5 + N_2)!}\right]\;,\label{UR_Gamma21}\\
 \Gamma^{(2)}_{20} =& \frac{\beta^2}{12} \left[1 - \frac{24 (2 + N_2)!}{(5 + N_2)!}\right]\;,\label{UR_Gamma22}
\end{align}
which again reduce  
to the basis-free result~(\ref{UR_R1neg}), (\ref{UR_R2neg}) in the limit $N_1,N_2 \rightarrow \infty$. 

\begin{table*}
	\renewcommand*{\arraystretch}{2.5}
	\begin{tabular}{|r||c|c|c|c|c|}
		\hline
		& 
		$\ell_{V\pi}[\tau_V] = \tau_{V\pi} [\tau_V]$ & 
		$\lambda_{V\pi} [\tau_V]$ & 
		$\delta_{V\pi E} [\tau_V]$ & 
		$\delta_{VB} [\tau_V]$ & 
		$\delta_{\pi B} [\tau_\pi]$ \\\hline \hline
		$\mathcal{R}^{(\ell)}_{-r,0}$ & 
		$0$ & 
		$\beta/16$ & 
		$\beta^2/48$ & 
		$3 \beta/4$ & 
		$\beta/2$ \\
		$\gamma^{(\ell)}_{r0}$ & 
		${\displaystyle \frac{\beta}{20}(1-4N_2)}$ & 
		${\displaystyle \frac{\beta}{20}(1 + N_2)}$ &
		${\displaystyle -\frac{\beta^2}{40}N_2(1+N_2)}$ & 
		${\displaystyle \frac{\beta}{12}(1 + 4N_1)}$ & 
		${\displaystyle \frac{2\beta}{5}(1 + N_2)}$ \\
		$\Gamma^{(\ell)}_{r0}$ & 
		${\displaystyle \frac{6\beta (1+N_2)!}{(5+N_2)!}}$ & 
		${\displaystyle \frac{\beta}{16}\left[1 - \frac{24(1+N_2)!}{(5+N_2)!}\right]}$ & 
		${\displaystyle \frac{\beta^2}{48}\left[1 - \frac{24(1+N_2)!}{(4+N_2)!}\right]}$ & 
		${\displaystyle \frac{3\beta}{4}\left[1 - \frac{16/3}{(N_1+2)(N_1+3)}\right]}$ & 
		${\displaystyle \frac{\beta}{2}\left[1 - \frac{24(1+N_2)!}{(5+N_2)!}\right]}$ 
		\\[0.1cm]\hline
	\end{tabular}
	\caption{The transport coefficients $\ell_{V\pi}, \tau_{V\pi}, \lambda_{V\pi}$, $\delta_{\pi B}, \delta_{VB}, \delta_{V\pi E}$
		for an ultrarelativistic ideal gas. Their values are computed by inserting $\mathcal{R}^{(\ell)}_{-r,0}$ from 
		Eq.~\eqref{UR_Rr}, $\gamma^{(\ell)}_{r0}$ from Eqs.~(\ref{UR_gamma1}), (\ref{UR_gamma2}), and 
		$\Gamma^{(\ell)}_{r0}$ from Eqs.~(\ref{UR_Gamma11})--(\ref{UR_Gamma22}). The relation between 
		$\lambda_{V\pi}$ and $\ell_{V\pi}$ reported in Eq.~\eqref{eq:UR_ell_tau_lambda_V_pi} holds in all three cases. 
		The results obtained using $\Gamma^{(\ell)}_{r0}$ agree with those obtained using $\gamma^{(\ell)}_{r0}$ and 
		$\mathcal{R}^{(\ell)}_{-r,0}$ when $(N_1,N_2) = (1,0)$ and when $N_1,N_2 \rightarrow \infty$, respectively.
		\label{tbl:tcoeffs}
	}
\end{table*}

\begin{figure*}
	\begin{tabular}{cc}
		\includegraphics[width=1.0\columnwidth]{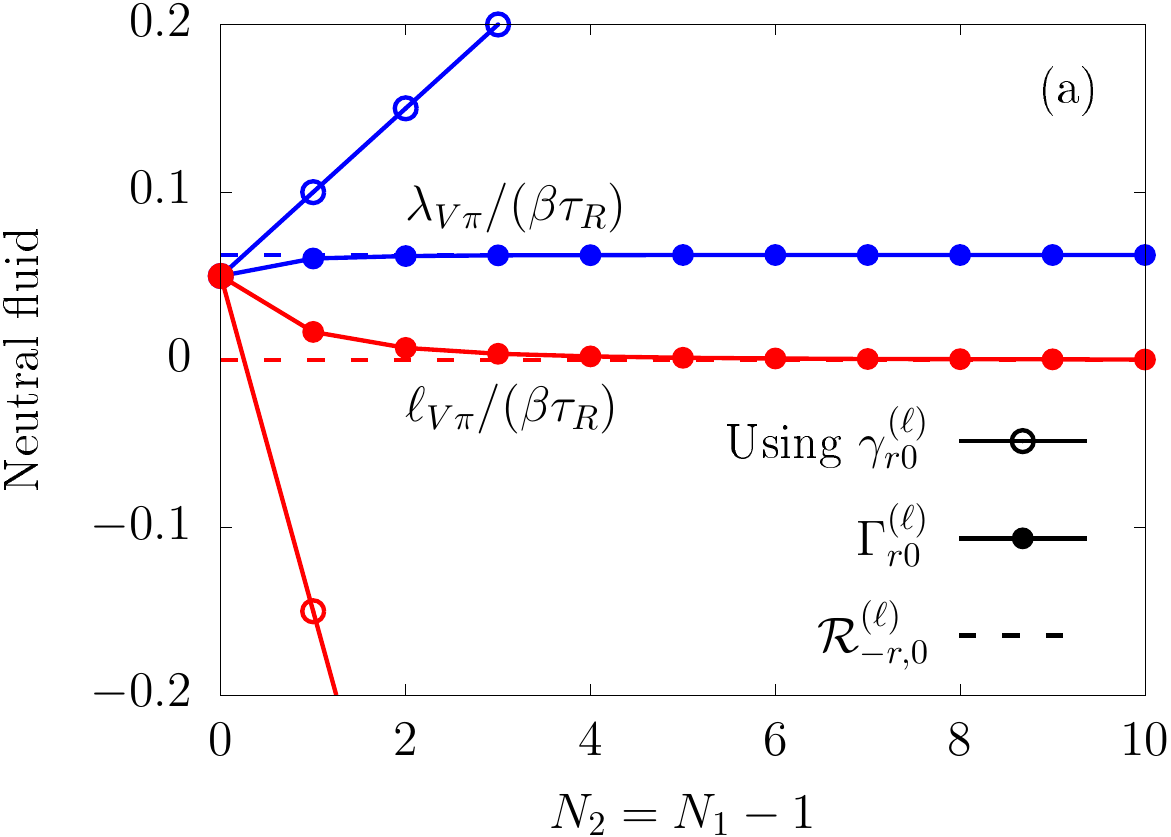} &
		\includegraphics[width=1.0\columnwidth]{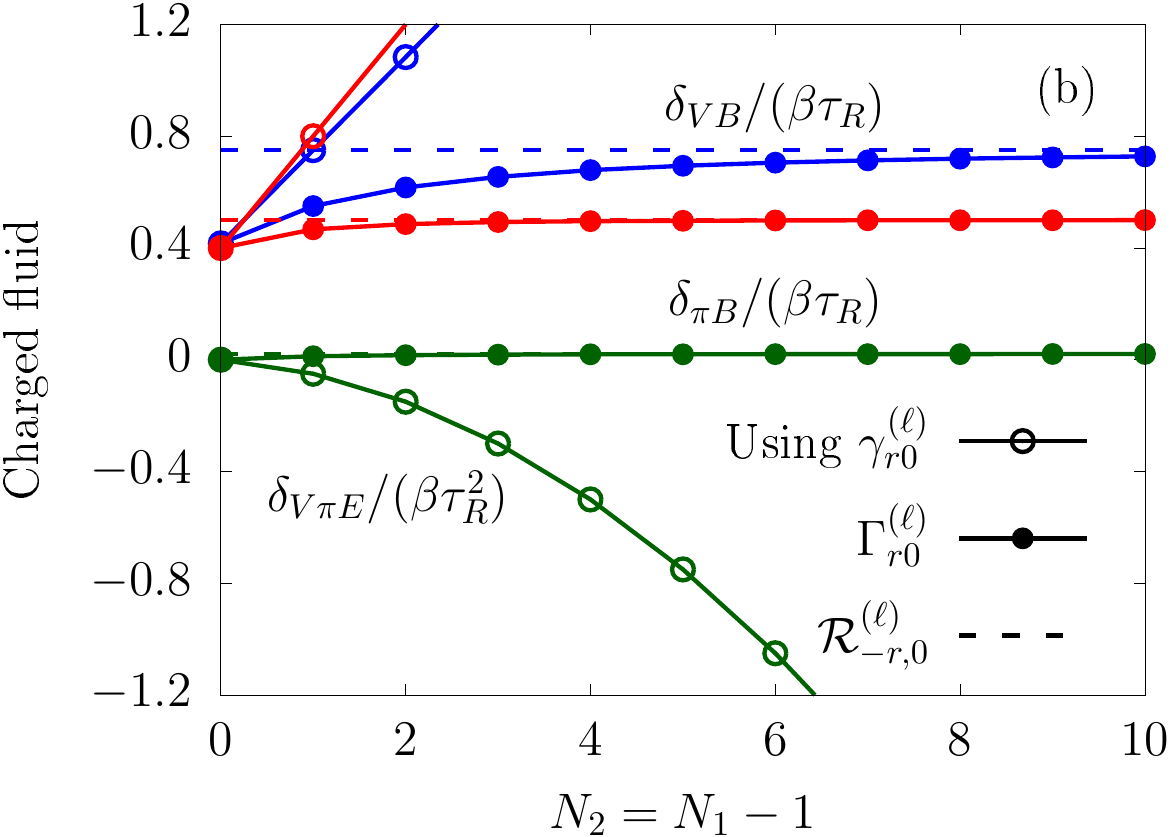}
	\end{tabular}
	\caption{Dependence on $N_2 = N_1 - 1$ of the coefficients (a)
		$\ell_{V\pi} = \tau_{V\pi}$, $\lambda_{V\pi}$ for a neutral fluid; 
		and (b) $\delta_{V\pi E}$, $\delta_{VB}$, $\delta_{\pi B}$ for a 
		charged fluid, computed using the approaches shown in 
		Table~\ref{tbl:tcoeffs}.\label{fig:lVpi}}.
\end{figure*}

\subsection{Transport coefficients for the ultrarelativistic ideal gas}\label{sec:UR:tcoeffs}

We now employ the basis-free results~\eqref{UR_R1neg},\eqref{UR_R2neg} 
for $\mathcal{R}^{(1)}_{-1,0}$, $\mathcal{R}^{(2)}_{-1,0}$, and 
$\mathcal{R}^{(2)}_{-2,0}$. The ultrarelativistic limit of the transport 
coefficients appearing in Eqs.~\eqref{bf_kappa}--\eqref{bf_lambda_V_pi} 
is then obtained as
\begin{align}
	\kappa &= \frac{\beta P}{12} \,\tau_V \;, \quad 
	\delta_{VV} = \tau_V\;, \quad 
	\lambda_{VV} = \frac{3}{5}\, \tau_V\;, \nonumber\\
	\ell_{V \pi} &= \tau_{V \pi} = 0\;, \qquad
	\lambda_{V\pi} = \frac{\beta}{16}\,  \tau_V\;.
	\label{eq:UR_vector}
\end{align}
Equations~(\ref{bf_eta})--(\ref{bf_lambda_pi_V}) reduce to
\begin{gather}
	\eta = \frac{4 P}{5} \, \tau_\pi\;, \quad 
	\delta_{\pi\pi} =  \frac{4}{3}\, \tau_\pi\;, \quad 
	\tau_{\pi\pi} = \frac{10}{7}\, \tau_\pi\;,\nonumber\\
	\ell_{\pi V} = \tau_{\pi V} = \lambda_{\pi V} = 0\;.
	\label{eq:UR_tensor}
\end{gather}
The coefficients in Eqs.~\eqref{bf_delta_V_E}-\eqref{bf_delta_pi_B} 
due to the electric and magnetic fields read 
\begin{align}
	\delta_{VE} &= \frac{\beta^2 P}{12}\, \tau_V\;, \quad 
	\delta_{\pi V E} = \frac{8}{5}\, \tau_\pi\;, \nonumber\\ 
	\delta_{\pi B} & = \beta \tau_\pi\;, \quad
	\delta_{VB} = \frac{3\beta}{4}\,  \tau_V\;, \qquad
	\delta_{V\pi E} = \frac{\beta^2 }{48}\, \tau_V\;.
	\label{eq:UR_MHD}
\end{align}
In the above, the coefficients involving the bulk viscous pressures were omitted.

For the ideal ultrarelativistic gas, Eq.~\eqref{UR_Rr} can be employed to show that 
\begin{equation}
 \frac{\partial \mathcal{R}^{(\ell)}_{r0}}{\partial \alpha} = 0\;, \qquad 
 \frac{\partial \mathcal{R}^{(\ell)}_{r0}}{\partial \beta} = -\frac{r}{\beta} \mathcal{R}^{(\ell)}_{r0}\;.
\end{equation}
The above relations hold true also when $r < 0$ and in particular 
also when $\mathcal{R}^{(\ell)}_{-r,0}$ is replaced by 
$\gamma^{(\ell)}_{r0}$ or $\Gamma^{(\ell)}_{r0}$, since their 
dependence on $\alpha$ and $\beta$ is identical to that of 
$\mathcal{R}^{(\ell)}_{-r,0}$. Thus, one can conclude that in all 
approaches mentioned here, 
\begin{equation}
 \tau_{V\pi} = \ell_{V\pi}\;, \qquad 
 \lambda_{V\pi} = \frac{\tau_V}{16}\beta - \frac{\ell_{V\pi}}{4}\;.\label{eq:UR_ell_tau_lambda_V_pi}
\end{equation}

Since the coefficients $\ell_{V\pi}$, $\tau_{V\pi}$, $\lambda_{V\pi}$,
$\delta_{\pi B}$, $\delta_{VB}$, and $\delta_{V\pi E}$  involve 
$\mathcal{R}^{(1)}_{-1,0}$, $\mathcal{R}^{(2)}_{-1,0}$, and 
$\mathcal{R}^{(2)}_{-2,0}$, their values will differ 
between the various approaches discussed in the present section. All other transport coefficients assume the
same values as in the standard DNMR approach.
As pointed out in Table~\ref{tbl:tcoeffs}, when $N_\ell \rightarrow \infty$, the 
approach based on $\Gamma^{(\ell)}_{r0}$ converges to the basis-free one employing 
$\mathcal{R}^{(\ell)}_{-r,0}$. Conversely, the coefficients computed based on 
$\gamma^{(\ell)}_{r0}$ diverge with the truncation order $N_\ell$. 
We illustrate these behaviours in Fig.~\ref{fig:lVpi} for the coefficients shown 
in Table~\ref{tbl:tcoeffs}. Note that in the 14-moment approximation, 
when $N_0 = 2$, $N_1 = 1$, and $N_2 = 0$, the results obtained using the coefficients 
$\gamma^{(\ell)}_{r0}$ and $\Gamma^{(\ell)}_{r0}$ are identical and reproduce those 
reported in Refs.~\cite{Denicol:2012cn,Denicol:2019iyh}.

\section{Shear-diffusion coupling: Longitudinal waves} \label{sec:long}

In this section, we consider the propagation of longitudinal (sound) 
waves through an ultrarelativistic, uncharged ideal fluid. The purpose 
of this section is to compare the prediction of second-order fluid dynamics 
using the various expressions of the transport coefficients reported in 
Table~\ref{tbl:tcoeffs} with that of kinetic theory in RTA. While the 
former can be estimated analytically, the latter is obtained numerically 
using the method described in Ref.~\cite{Ambrus:2017keg}. 
Per definition, a sound wave is an infinitesimal perturbation, such that it is sufficient to consider the linear terms in
the equations of motion. In the linearized equations of motion for an ultrarelativistic, uncharged fluid, only
the coefficients $\ell_{V \pi}, \ell_{\pi V}$ enter (as well as some coefficients in 
$\mathcal{K}^\mu$ and $\mathcal{K}^{\mu \nu}$, which, however, play no role in our investigation, see comment
after Eq.~\eqref{J_munu}).
Since $\ell_{\pi V}$ vanishes in all approaches considered here, we will refer only to the coefficient $\ell_{V\pi}$ 
listed in Table~\ref{tbl:tcoeffs}, for which we summarize the results below:
\begin{align}
 \text{Basis-free:}& & 
 \ell_{V\pi} &= 0\;,\label{lVpi_bf}\\
 \text{DNMR:}& & 
 \ell_{V\pi} &= \frac{\beta}{20}(1 - 4N_2) \tau_V\;, \label{lVpi_DNMR}\\
 \text{Corrected DNMR:}& & 
 \ell_{V\pi} &= \frac{6 \beta(N_2 + 1)!}{(N_2 + 5)!}\,\tau_V\;. \label{lVpi_IReD}
\end{align}
In addition, we recall the result reported in Ref.~\cite{Panda:2020zhr}, 
obtained using a second-order Chapman-Enskog approach:
\begin{equation}
 \text{Ref.~\cite{Panda:2020zhr}:} \qquad
 \ell_{V\pi} = \frac{\beta}{4}\, \tau_V\;.
 \label{lVpi_CE}
\end{equation}
We note that the result $\ell_{V\pi} = 0$ was also obtained in Ref.~\cite{Jaiswal:2015mxa}
using a Chapman-Enskog--like approach.

Since the corrected DNMR value lies between the DNMR 
(for $N_2 = 0$) and basis-free (for $N_2 \rightarrow \infty$) results, 
we will not consider it explicitly in what follows. Instead, we will 
contrast the basis-free prediction to predictions due to 
Ref.~\cite{Panda:2020zhr} and to the DNMR prediction, where for illustrative purposes
we choose $N_2 = 2$, leading to $\ell_{V\pi} = -7 \beta \tau_V / 20$. 

This section is structured as follows. In Sec.~\ref{sec:long:eqs}, 
we derive the equations of motion for sound waves. 
The resulting dispersion relations are computed in Sec.~\ref{sec:long:modes}. 
The analytical solutions and the numerical results are discussed 
in Sec.~\ref{sec:long:res}.

\subsection{Second-order equations for longitudinal waves}\label{sec:long:eqs}

We assume that the background fluid is 
homogeneous and at rest, while the perturbations travel along 
the $z$ axis. The velocity of the perturbed fluid is
$u^\mu = \gamma (1, 0, 0, \delta v)\simeq (1,0,0,\delta v)$, 
where $|\delta v| \ll 1$ is assumed to be small. For simplicity, the transverse motion 
leading to so-called shear waves is not taken 
into account. The properties of the background fluid are 
\begin{equation}
 e = e_0 + \delta e\;, \qquad 
 n = n_0 + \delta n\;,
\end{equation}
where again $|\delta e| / e_0, 
|\delta n|/ n_0 \ll 1$. The diffusion vector $V^\mu$ and 
shear-stress tensor $\pi^{\mu\nu}$ can be described 
in terms of only two scalar quantities, $\delta V$ and $\delta \pi$, as follows:
\begin{align}
 V^\mu =& \delta V(\delta v, 0, 0, 1)\;, 
\end{align}
and
\begin{align}
 \pi^{\mu\nu} =& \delta \pi
 \begin{pmatrix}
  \delta v^2 \gamma^2 & 0 & 0 & \delta v \gamma^2 \\
  0 & -\frac{1}{2} & 0 & 0 \\
  0 & 0 & -\frac{1}{2} & 0 \\
  \delta v \gamma^2 & 0 & 0 & \gamma^2
 \end{pmatrix}\;,
 \label{eq:long_npi}
\end{align}
where the properties $u_\mu V^\mu = u_\mu \pi^{\mu\nu} = \pi^\mu{}_\mu = 0$ were employed.
Since both $\delta V$ and $\delta \pi$ are related to gradients 
of the fluid, they are of the same order of magnitude
as the perturbations. In the linearized limit, $V^\mu$ and $\pi^{\mu\nu}$ reduce to
\begin{equation}
\hspace{-5pt} V^\mu \simeq \delta V (0, 0, 0, 1)\;, \ 
 \pi^{\mu\nu} \simeq \delta \pi \ {\rm diag}\left(0,-\frac{1}{2}, 
 -\frac{1}{2}, 1\right)\;.
\end{equation}

Noting that the expansion scalar 
$\theta$ and the shear tensor $\sigma^{\mu\nu}$ reduce to
\begin{equation}
 \theta = \partial_z \delta v\;, \qquad 
 \sigma^{\mu\nu} = {\rm diag}\left(0,\frac{1}{3}, \frac{1}{3}, 
 -\frac{2}{3}\right) \partial_z \delta v\;,
\end{equation}
while 
\begin{equation}
 \Delta^\lambda_\mu \nabla_\nu \pi^{\mu\nu} = 
 \partial_z \pi^{\lambda z} = \delta^\lambda_z \partial_z \delta \pi\;,
 \label{eq:long_Delta_nabla_pi}
\end{equation}
the conservation equations \eqref{eq:cons_n}--\eqref{eq:cons_u} become
\begin{align}
 \partial_t \delta n + n_0 \partial_z \delta v + \partial_z \delta V =& 0\;,\nonumber\\
 \partial_t \delta e + (e_0 + P_0) \partial_z \delta v =& 0\;,\nonumber\\
 (e_0 + P_0)\partial_t \delta v + \partial_z \delta P + 
 \partial_z \delta \pi =& 0\;.
 \label{eq:long_SET}
\end{align}

The equations of motion for $\delta V$ and $\delta \pi$ can be obtained from 
Eqs.~(\ref{Vdot}),(\ref{pidot}) and (\ref{J_mu}),(\ref{J_munu}) 
by ignoring terms that are quadratic with respect to the perturbations:
\begin{align}
 \tau_V \dot{V}^{\langle \mu\rangle} + V^\mu =& \kappa \nabla^\mu\alpha + 
 \ell_{V\pi} \Delta^{\mu\nu} \nabla_\lambda \pi^{\lambda}_\nu\;,\nonumber\\
 \tau_{\pi} \dot{\pi}^{\langle \mu\nu \rangle} + \pi^{\mu\nu} =& 
 2\eta \sigma^{\mu\nu}
 + \ell_{\pi V} \nabla^{\langle \mu} 
 V^{\nu \rangle}\;.
\end{align}
Using $\dot{V}^{\langle \mu \rangle} \simeq \delta^\mu_z \partial_t \delta V$, 
$\dot{\pi}^{\langle zz \rangle} \simeq \partial_t \delta \pi$ and 
noting that $\ell_{\pi V} = 0$ by virtue of Eq.~\eqref{eq:UR_tensor}, we find
\begin{align}
 \tau_V \partial_t \delta V + \delta V =& -\kappa \partial_z \delta \alpha  +
 \ell_{V\pi} \partial_z \delta \pi\;,\nonumber\\
 \tau_\pi \partial_t \delta \pi + \delta \pi =& -\frac{4\eta}{3} \partial_z \delta v\;,
 \label{eq:long_diss}
\end{align}
where $\delta \alpha = \frac{4}{n_0} \delta n - \frac{3}{P_0} \delta P$. 
We shall employ the Knudsen number ${\rm Kn} \sim |k \tau_V|, |k \tau_\pi| \ll 1$ 
for power-counting purposes in order to simplify some of the expressions 
appearing in the following sections.

\subsection{Mode analysis}\label{sec:long:modes}

Now we perform the analysis of Eqs.~\eqref{eq:long_SET} and \eqref{eq:long_diss} at the level of the Fourier 
modes corresponding to $e^{-i (\omega t - k z)}$, introduced for a quantity $A(t, x)$ as
\begin{equation}
 A(t,x) = A_0 + 
 \int_{-\infty}^\infty dk
 \sum_\omega e^{-i(\omega t - k z)} 
 \delta A_\omega(k)\;,
\end{equation}
where $A_0$ is the constant background value of $A$, while $|k| = 2\pi / \lambda$ is the wavenumber 
(not to be confused with the particle momentum $k^\mu$ from the previous sections) and 
$\omega \equiv \omega(k)$ is the angular 
frequency, whose real part gives rise to propagation. A negative imaginary part of $\omega$ leads to damping of the 
mode. A positive imaginary part would lead to an exponential increase and thus to an instability.
Applying the above Fourier expansion leads to the matrix equation
\begin{multline}
 \hspace*{-0.4cm} \begin{pmatrix}
  -3\frac{\omega}{k} & 4P_0 & 0 & 0 & 0\\
  1 & -\frac{4\omega}{k} P_0 & 1 & 0 & 0\\
  0 & \frac{4\eta}{3} & -\frac{i}{k} - \frac{\omega}{k} \tau_\pi & 0 & 0\\
  0 & n_0 & 0 & -\frac{\omega}{k} & 1 \\
  -\frac{3\kappa}{P_0} & 0 & -\ell_{V\pi} & 
  \frac{4\kappa}{n_0} & -\frac{i}{k} - \frac{\omega}{k} \tau_V
 \end{pmatrix}
 \begin{pmatrix}
  \delta P_\omega(k) \\ \delta v_\omega(k) \\ \delta \pi_\omega(k) \\ 
  \delta n_\omega(k) \\ \delta V_\omega(k)
 \end{pmatrix} \\
 = 0\;.
\end{multline}
The modes supported by this system can be found by setting 
the determinant of the above matrix to $0$.
Since $\ell_{\pi V} = 0$, the $(\delta P, \delta v, \delta \pi)$ sector 
decouples from the $(\delta n, \delta V)$ sector and the determinant factorizes as
\begin{align}
 (k^2 - 3 \omega^2) (1 - i \omega \tau_\pi) 
 - \frac{i k^2 \omega}{P_0}\eta =& 0\;,\label{factorization1}\\
 \omega(1 - i \omega \tau_V) + \frac{4i k^2}{n_0} \kappa =& 0\;. \label{factorization2}
\end{align}
The $(\delta P, \delta v, \delta \pi)$ sector contains the two sound or acoustic modes as well as a shear mode,
while the $(\delta n, \delta V)$ sector contains a mode associated with particle-number transport (in the
non-relativistic context called thermal mode) and a diffusive mode. While the sound modes and the thermal mode are
hydrodynamic modes (i.e., the frequency vanishes for zero wavenumber), 
the shear and the diffusive modes are non-hydrodynamic modes (i.e., the frequency does not vanish for
zero wavenumber).

Equations~\eqref{factorization1} and \eqref{factorization2} 
agree with Eqs.~(4.19) and (4.13) of Ref.~\cite{Ambrus:2017keg} when identifying 
$\omega = -i\alpha$ and $\kappa = \lambda/16$. Therefore, the dispersion relations $\omega \equiv \omega(k)$ are 
identical to those identified in Eqs.~(4.14) and (4.20)--(4.22) of Ref.~\cite{Ambrus:2017keg}. Labeling the 
acoustic and shear modes as $\omega_a^\pm$ and $\omega_\eta$, respectively, we have
\begin{equation} \label{eq:long_omega_eta}
 \omega^\pm_a = \pm |k| c_{s;a} -i \xi_a, \qquad 
 \omega_\eta = -i \xi_\eta\;, 
 \end{equation}
where the argument $k$ was omitted for brevity. The quantities appearing above are defined as
\begin{align}
 c_{s;a} =& \frac{1}{2 |k| \tau_\pi \sqrt{3}} 
 \left\{\frac{1}{R_\eta}\left[1 - k^2 \tau_\pi^2 \left(1 + \frac{\eta}{\tau_\pi P_0}\right)\right] - R_\eta\right\}\;,\nonumber\\
 \xi_a =& \frac{1}{3\tau_\pi} \left\{1 - 
 \frac{1}{2R_\eta} \left[1 - k^2 \tau_\pi^2 \left(1 + \frac{\eta}{\tau_\pi P_0}\right)\right] - \frac{R_\eta}{2}\right\}\;,\nonumber\\
 \xi_{\eta} =& \frac{1}{3\tau_\pi} \left\{1 +
 \frac{1}{R_\eta} \left[1 - k^2 \tau_\pi^2 \left(1 + \frac{\eta}{\tau_\pi P_0}\right)\right] + R_\eta\right\}\;.
\end{align}
Here, the function $R_\eta$ is defined as
\begin{align}
 R_\eta =& 
 \begin{cases}
  R_\eta^<, & \tau_\pi < \tau_{\pi,{\rm lim}}\;,\\
  -R_\eta^>, & \tau_\pi > \tau_{\pi;{\rm lim}}\;,
 \end{cases}\nonumber\\
 R_\eta^< =& \left[1 - 3|k| \tau_\pi \sqrt{R_{\eta,{\rm aux}}} 
 + 3 k^2 \tau_\pi^2 \left(1 - \frac{\eta}{2P_0 \tau_\pi}\right)\right]^{1/3}\;,\nonumber\\
 R_\eta^> =& \left[-1 + 3 |k|\tau_\pi \sqrt{R_{\eta,{\rm aux}}} 
 - 3k^2 \tau_\pi^2\left(1 - \frac{\eta}{2P_0 \tau_\pi}\right)\right]^{1/3}\;,
\end{align}
with
\begin{multline}
 R_{\eta,{\rm aux}} = 1 + \frac{2}{3} k^2 \tau_\pi^2 
 \left(1 - \frac{5\eta}{2P_0 \tau_\pi} - \frac{\eta^2}{8P_0^2 \tau_\pi^2}\right) \\
 + \frac{k^4 \tau_\pi^4}{9} \left(1 + \frac{\eta}{P_0 \tau_\pi}\right)^3\;.
\end{multline}
In the above, the value $\tau_{\pi, {\rm lim}}$ discerning between the two branches for $R_\eta$ is given by
\begin{equation}
 \tau_{\pi, {\rm lim}} = \frac{1}{|k|} 
 \left(1 + \frac{\eta}{P_0 \tau_\pi}\right)^{-1/2}\;,
\end{equation}
where $\eta / (P_0 \tau_\pi)$ is independent of $\tau_\pi$ since $\eta \sim \tau_\pi$.
Applying the power-counting scheme mentioned above, we observe that 
$c_{s;a} \simeq c_s + O({\rm Kn}^2)$, with $c_s = 1/\sqrt{3}$ the speed of sound, while
$\xi_a \simeq \frac{k^2 \eta}{6P_0} + O({\rm Kn}^3)$, and
$\xi_\eta \simeq \frac{1}{\tau_\pi} - \frac{k^2 \eta}{3P_0} + O({\rm Kn}^3)$.

The thermal and diffusive modes, $\omega^-_\kappa$ and $\omega^+_\kappa$,
respectively, are 
\begin{align}
 \omega^\pm_\kappa = -i \xi^\pm_\kappa\;,\, \,
 \xi^\pm_\kappa = \frac{1}{2 \tau_V} 
 \left(1 \pm \sqrt{1 - \frac{16 k^2 \kappa \tau_V}{n_0}}\right)\;,
 \label{eq:long_omega_kappa}
\end{align}
and agree with Eq.~(4.14) of Ref.~\cite{Ambrus:2017keg}. A power-counting analysis reveals that
$\xi^-_\kappa \simeq \frac{4k^2 \kappa}{n_0} + O({\rm Kn}^3)$
and $\xi^+_\kappa \simeq \frac{1}{\tau_V} - \frac{4k^2 \kappa}{n_0} + O({\rm Kn}^3)$.

With the dispersion relations at hand, we can now compute the mode amplitudes.
Focusing first on the thermal and diffusive modes, it is not difficult to see 
that $\delta P^\pm_{\kappa}(k) = \delta v^\pm_{\kappa}(k) = \delta \pi^{\pm}_{\kappa} (k)= 0$, 
while the amplitude of the diffusion current can be linked to 
that of the density fluctuations via
\begin{equation}
 \delta V^{\pm}_{\kappa}(k) = -\frac{i \xi^\pm_\kappa}{k} \delta n^\pm_\kappa(k)\;.
\end{equation}
In the sound and shear sector, the amplitude of the pressure fluctuations can 
be defined as an independent variable, while the other amplitudes 
can be expressed as
\begin{align}
 \delta v_\omega(k) &= \frac{3\omega}{4kP_0} 
 \delta P_\omega(k)\;, \quad
 \delta \pi_{\omega} (k)= \left(\frac{3\omega^2}{k^2} - 1\right) \delta P_{\omega} (k)\;,\nonumber\\
 \delta n_{\omega} (k) &= \left[\frac{3n_0}{4P_0} + 
 \frac{i \ell_{V\pi} (3\omega^2 -k^2)}
 {\frac{4i k^2 \kappa}{n_0} + 
 \omega(1 - i \omega \tau_V)}
 \right]\delta P_{\omega} (k)\;,\nonumber\\
 \delta V_{\omega} (k)&= \frac{i\omega}{k} \frac{\ell_{V\pi}(3\omega^2 -k^2)}
 {\frac{4i k^2 \kappa}{n_0} + 
 \omega(1 - i \omega \tau_V)} \delta P_{\omega} (k)\;,
\end{align}
where $\omega$ is either $\omega_a^\pm$ or $\omega_\eta$.
From the above, it is clear that a non-vanishing value of $\ell_{V\pi}$ 
introduces acoustic and shear modes into the diffusion current, allowing the diffusion 
current to propagate by means of the sound modes. Thus, the basis-free 
result $\ell_{V\pi} = 0$ can be distinguished from the Chapman-Enskog and 
DNMR results $\ell_{V\pi} \neq 0$ by considering the propagation of a simple 
harmonic wave, which we discuss below.

\subsection{Numerical results}\label{sec:long:res}

At initial time $t_0 = 0$, we consider
\begin{equation}
 n(t_0, z) = n_0\;, \ 
 P(t_0, z) = P_0 + \delta P \cos(k z)\;,
 \label{eq:long_ic}
\end{equation}
while $\delta v(t_0, x) = \delta \pi(t_0, z) = \delta V(t_0, z) = 0$. This initial state can be implemented by setting 
\begin{equation}
 \delta P_\omega(k') = \frac{\delta P_\omega}{2} 
 [\delta(k' - k) + \delta(k' + k)],
\end{equation}
with $\sum_\omega \delta P_\omega = \delta P$. 
This allows the solutions for 
$\delta P(t,z)$, $v(t,z)$, and $\delta \pi(t,z)$ to be written as
\begin{align}
 \delta P(t,z) =& \cos(kz) \sum_{\omega^\pm_a, \omega_\eta} \delta P_\omega e^{-i \omega t}\;,\nonumber\\
 \delta v(t,z) =& \frac{3i}{4 k P_0} \sin(kz) 
 \sum_{\omega^\pm_a, \omega_\eta} \omega \delta P_\omega e^{-i \omega t}\;,\nonumber\\
 \delta \pi(t,z) =& \cos(kz) \sum_{\omega^\pm_a, \omega_\eta} \left(\frac{3\omega^2}{k^2} - 1\right) 
 \delta P_\omega e^{-i \omega t}\;. 
\end{align}
Imposing the initial conditions from Eq.~\eqref{eq:long_ic} leads to
\begin{gather}
 \sum_{\omega^\pm_a, \omega_\eta}\delta P_\omega = \delta P\;,\quad
 \sum_{\omega^\pm_a, \omega_\eta} \omega \delta P_\omega = 0\;, \nonumber \\
 \sum_{\omega^\pm_a, \omega_\eta} \omega^2 \delta P_\omega = \frac{k^2}{3} \delta P\;,
\end{gather}
which admits the solutions
\begin{align}
 \delta P_{a}^\pm =& \pm \frac{k^2 + 3 \omega_\eta \omega_a^\mp}{6|k|c_{s;a} 
 (\omega_a^\pm - \omega_\eta)} \delta P \;,\nonumber\\
 \delta P_\eta =& -\frac{k^2(1 - 3c_{s;a}^2) - 3 \xi_a^2}
 {3[k^2 c_{s;a}^2 + (\xi_a - \xi_\eta)^2]} \delta P\;.
\end{align}
For small ${\rm Kn}$, we have 
\begin{align}
 \delta P_{a}^\pm =& \frac{\delta P}{2} \pm \frac{i|k| \eta c_s}{4 P_0} \delta P + O({\rm Kn}^3)\;, \nonumber\\
 \delta P_\eta =& \frac{k^4 \eta \tau_\pi^3}{9P_0} \delta P + O({\rm Kn}^6)\;.
\end{align}

\begin{figure}
\includegraphics[width=0.95\linewidth]{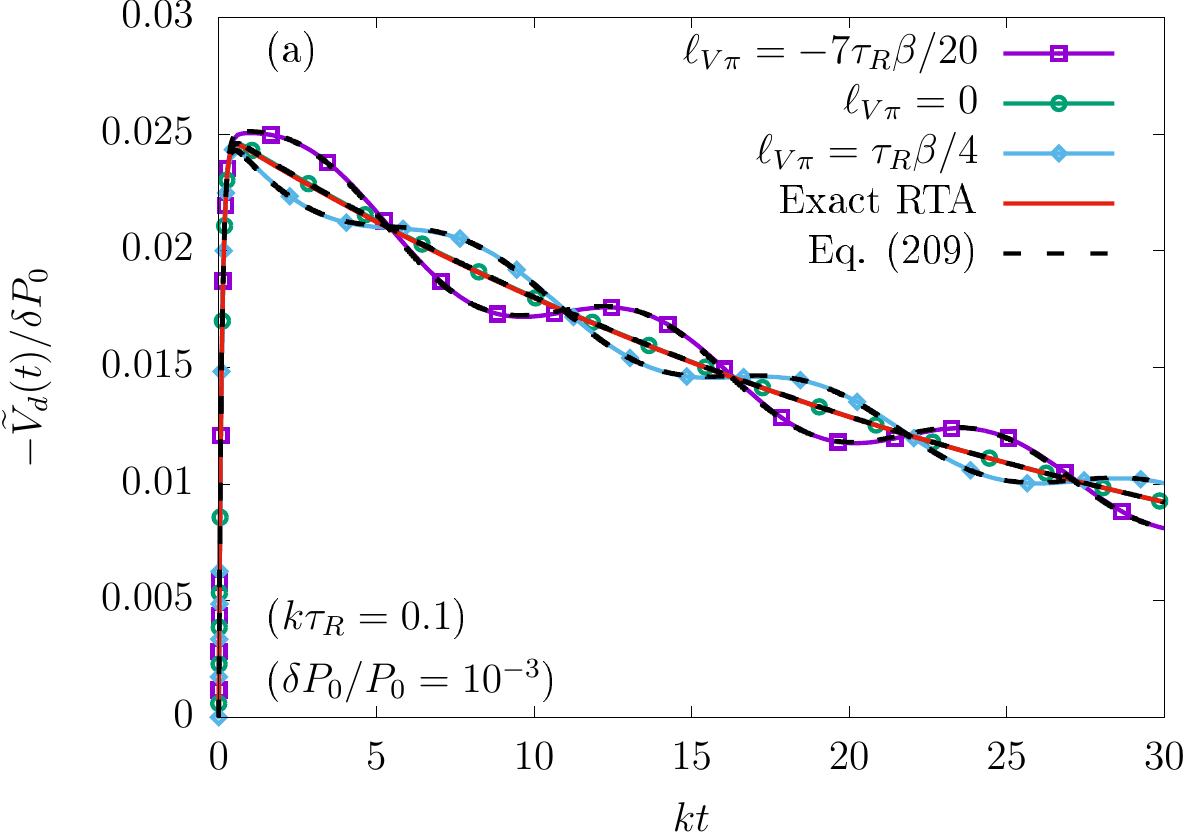}\\
\includegraphics[width=0.95\linewidth]{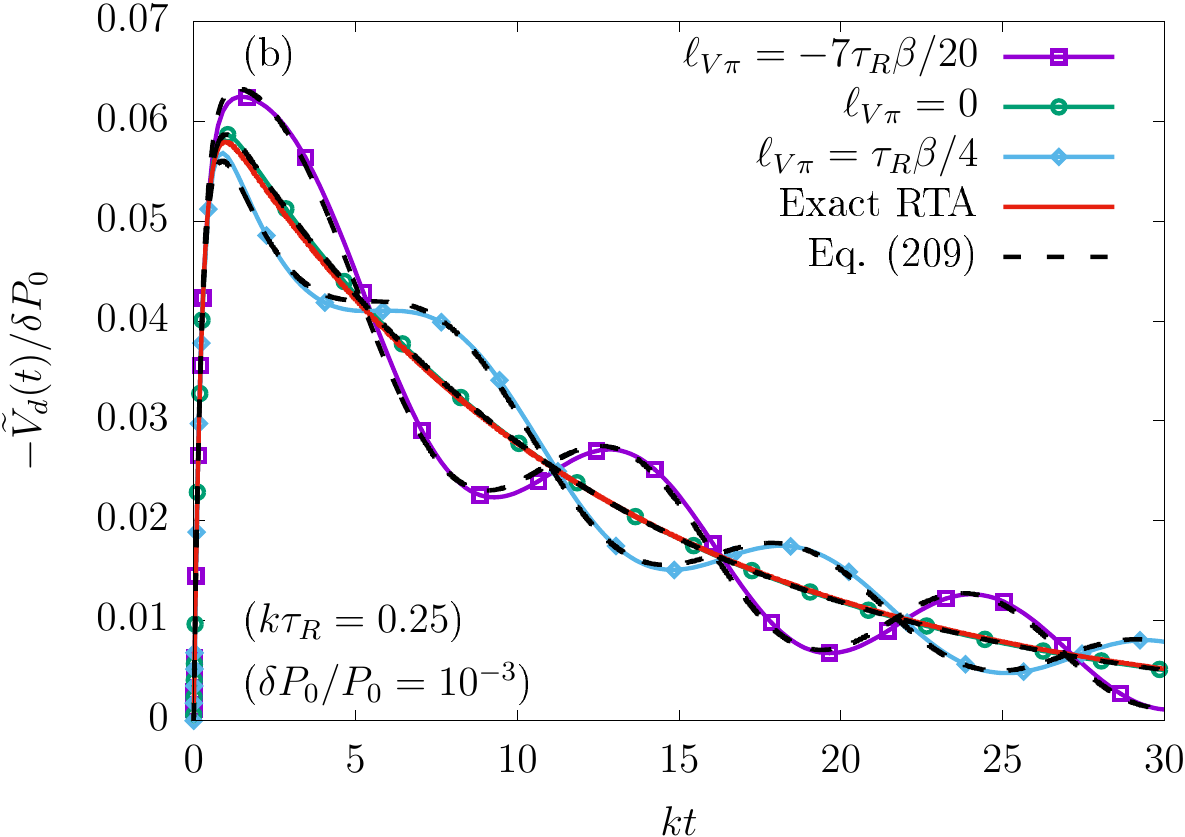}
\caption{
Time evolution of $-\widetilde{V}_d(t) / \delta P_0$ for 
the initial conditions in Eq.~\eqref{eq:long_ic}. 
The numerical solutions of the linearized equations 
\eqref{eq:long_SET} and \eqref{eq:long_diss} are shown 
using lines and symbols for various values of $\ell_{V\pi}$. 
The dashed black lines show their approximate analytical 
solution given in Eq.~\eqref{eq:long_nd_sol}. The numerical 
solution of the Boltzmann equation in RTA is shown with the solid red line. 
All results are obtained for $k \tau_R = 0.1$ (a) and $0.25$ 
(b) and we considered $\delta P_0 / P_0 = 10^{-3}$.
\label{fig:long}}
\end{figure}

To correctly assess the role of $\ell_{V\pi}$, we first note that for the shear mode,
the factor $1 - i \omega_\eta \tau_V \simeq 1 - \frac{\tau_V}{\tau_\pi} + O({\rm Kn}^2)$.
For $\tau_V = \tau_\pi = \tau_R$, this is of order $O({\rm Kn}^2)$, while it is of
order $O({\rm Kn}^0)$ when $\tau_V \neq \tau_\pi$. 
Focusing now on the particle-number fluctuations, we may write 
$\delta n_\omega(k') = \frac{\delta n_\omega}{2} [\delta(k' - k) + \delta(k' + k)]$, 
where the amplitude of the corresponding acoustic and shear modes are obtained up to second order 
in ${\rm Kn}$ as
\begin{align}
 \delta n_{a}^\pm &\simeq \frac{n_0}{2P_0} \left(\frac{3}{4} \pm 
 \frac{3 i |k| c_s}{8P_0} \eta + \frac{k^2}{n_0} \ell_{V\pi} \eta\right) \delta P\;,\nonumber\\
 \delta n_{\eta} &\simeq \frac{k^2 \tau_R n_0}{n_0 \eta - 12 P_0 \kappa} \ell_{V\pi} \eta \delta P\;.
\end{align}
For the diffusion current, we write 
$\delta V_\omega(k') = -\frac{i \delta V_\omega}{2} [\delta(k' - k) - \delta(k' + k)]$,
where
\begin{align}
 \delta V_a^{\pm} &\simeq \pm \frac{i k^3 c_s \delta P}{2 P_0 |k|} \ell_{V\pi} \eta \;, & 
 \delta V_\eta &\simeq \frac{k n_0 \delta P}{n_0 \eta - 12 P_0 \kappa} \ell_{V\pi} \eta\;.
\end{align}
The amplitudes of the thermal and diffusive modes
$\delta n^\pm_{\kappa}$ can be found by noting that 
\begin{align}
 \delta n(t_0, z) \simeq& \cos(kz)\left[
 \frac{3n_0 \delta P}{4P_0} + \delta n^+_\kappa + \delta n^-_\kappa\right.\nonumber\\
 & \left. + \left(1 + \frac{\tau_R n_0 P_0}{n_0 \eta - 12 P_0 \kappa}\right)
 \frac{k^2}{P_0} \ell_{V\pi} \eta \delta P\right]\;,\nonumber\\
 \delta V(t_0, z) \simeq& \frac{\sin(kz)}{k}\left(
 \frac{k^2 n_0 \ell_{V\pi} \eta \delta P}{n_0 \eta - 12 P_0 \kappa} + 
 \xi^+_\kappa \delta n^+_\kappa + \xi^-_\kappa \delta n^-_\kappa\right)\;,
\end{align}
where only terms up to second order with respect to ${\rm Kn}$ were shown. 
Imposing $\delta n(t_0,z) = V_d(t_0,z) = 0$ gives
\begin{multline}
 \delta n_\kappa^\pm \simeq \pm \frac{n_0 \xi^{\mp} \delta P}
 {P_0(\xi^+_\kappa - \xi^-_\kappa)} 
 \left(\frac{3}{4} + \frac{\tau_R k^2 P_0 \ell_{V\pi} \eta}{n_0 \eta - 12 P_0 \kappa} + \frac{k^2}{n_0} \ell_{V\pi} \eta\right) \\
 \mp \frac{k^2 n_0}{(n_0 \eta - 12 P_0 \kappa)(\xi^+_\kappa - \xi^-_{\kappa})} \ell_{V\pi} \eta \delta P\;,
\end{multline}
while $\delta V^\pm_{\kappa} = \xi^\pm_\kappa \delta n^\pm_\kappa / k$. Noting that 
\begin{equation}
 \xi^+_\kappa \xi^-_\kappa = \frac{4 k^2 \kappa}{n_0 \tau_R}\;, \quad 
 \xi^+_\kappa - \xi^-_\kappa = \frac{1}{\tau_R} \sqrt{1 - \frac{16 k^2 \kappa \tau_R}{n_0}}\;,
\end{equation}
we obtain $\delta V(t,z)$ as
\begin{multline}
 \delta V(t,z) \simeq \frac{k n_0 \delta P}{P_0} \sin(k z)
 \Bigg[
 \frac{|k|}{n_0} c_s \ell_{V\pi} \eta e^{-\xi_a t} \sin(k c_{s;a} t) \\
 + \frac{P_0 \ell_{V\pi} \eta}{n_0  \eta - 12 P_0 \kappa} \left(e^{-\xi_\eta t} - 
 \tau_R \frac{\xi^+_\kappa e^{-\xi^+_\kappa t} - \xi^-_\kappa e^{-\xi^-_\kappa t}}
 {\sqrt{1 - \frac{16k^2 \kappa}{n_0}\tau_R}}\right)\\
 + \frac{\kappa}{n_0} \left(3 + \frac{4 \tau_R k^2 P_0 \ell_{V\pi} \eta}{n_0 \eta - 12 P_0 \kappa} 
 + \frac{4k^2}{n_0} \ell_{V\pi} \eta\right) 
 \frac{e^{-\xi^+_\kappa t} - e^{-\xi^-_\kappa t}}
 {\sqrt{1 - \frac{16k^2 \kappa}{n_0}\tau_R}}\Bigg]\;.
 \label{eq:long_nd_sol}
\end{multline}
It can be seen that $\ell_{V\pi}$ introduces an oscillatory piece in the diffusion current. In order to facilitate the analysis, 
we introduce the amplitudes $\widetilde{\delta e}$, $\widetilde{\delta v}$, $\widetilde{\delta \pi}$, $\widetilde{\delta n}$, and 
$\widetilde{\delta V}$ via 
\begin{align}
 \begin{pmatrix}
  \widetilde{\delta e}(t) \\
  \widetilde{\delta \pi}(t) \\
  \widetilde{\delta n}(t)
 \end{pmatrix} =& \frac{k}{\pi} \int_0^{2\pi/k} dz
 \begin{pmatrix}
  \delta e(t, z) \\ 
  \delta \pi(t, z) \\ 
  \delta n(t, z)
 \end{pmatrix} \cos(k z)\;,\\
 \begin{pmatrix}
  \widetilde{\delta v}(t) \\
  \widetilde{\delta V}(t) 
 \end{pmatrix} =& \frac{k}{\pi} \int_0^{2\pi/k} dz 
 \begin{pmatrix}
  \delta v(t, z) \\
  \delta V(t, z)
 \end{pmatrix}
 \sin(k z)\;.
 \label{eq:long_tilde}
\end{align}
The linearized equations \eqref{eq:long_SET} and \eqref{eq:long_diss} are then solved as a set of ODEs by replacing 
\begin{equation}
 \partial_z \left(\delta e, \delta v, \delta \pi, \delta n, \delta V\right) \rightarrow 
 k \left(-\widetilde{\delta e}, \widetilde{\delta v}, -\widetilde{\delta \pi}, -\widetilde{\delta n}, \widetilde{\delta V}\right)\;.
\end{equation}
Figure~\ref{fig:long} shows the results obtained using 
the values of $\ell_{V\pi} = -7\beta \tau_R / 20$, $0$, 
and $\beta \tau_R / 4$, as given by the DNMR approach 
based on $\gamma^{(2)}_1$ with $N_2 = 2$ \eqref{lVpi_DNMR}, 
the basis-free approach \eqref{lVpi_bf}, and in Ref.~\cite{Panda:2020zhr}, respectively.
The numerical results are compared with the analytical 
prediction \eqref{eq:long_nd_sol}, shown with dashed 
black lines. The small discrepancies seen in panel (b) 
are due to the 
approximations made in deriving Eq.~\eqref{eq:long_nd_sol}. 
Additionally, we also show with the solid red line the 
numerical solution of the Boltzmann equation \eqref{BTE} 
with the Anderson-Witting collision model \eqref{AW}, 
obtained as described in Ref.~\cite{Ambrus:2017keg}. 
The basis-free and RTA results are in excellent agreement, 
confirming that for the RTA, $\ell_{V\pi} = 0$.

\section{Conclusions}\label{sec:conc}

In this paper, we computed the transport coefficients of 
second-order relativistic fluid dynamics from the relativistic 
Boltzmann equation in the relaxation-time approximation (RTA) of the collision term. 

Employing the method of moments, the irreducible moments for a negative power of energy, the so-called
negative-order moments, are usually expressed in terms of the ones with a non-negative power of energy
using a kind of completeness relation, which becomes exact 
in the limit when the truncation order $N_\ell \rightarrow \infty$. Focusing on the 
14-dynamical moments approximation, we then considered different approaches to relate the negative-order moments 
$\rho_{-r}^{\mu_1 \cdots \mu_\ell}$ to the zeroth-order ones: (i) the original DNMR approach~\cite{Denicol:2012cn},
which features the coefficients  $\gamma^{(\ell)}_{r0}$, cf.\ Eq.~\eqref{gamma}, 
(ii) a corrected DNMR approach~\cite{Wagner:2022ayd},
which employs the coefficients $\Gamma^{(\ell)}_{r0}$ of Eq.~\eqref{Gamma}, (iii) a
so-called shifted-basis approach, which includes a certain set of negative-order moments in the expansion basis,
cf.\ Eq.~\eqref{shifted_Gamma}, and (iv) a basis-free approach tailored to the RTA, cf.\ Eq.~\eqref{bf_matching}.

The shifted-basis approach acknowledges the importance of the negative-order moments 
by including them explicitly in the expansion basis. The magnitude of the shifts 
$s_\ell$ for the irreducible moments of tensor rank $\ell$ are defined by the lowest-order moment 
$\rho^{\mu_1 \cdots \mu_\ell}_{-s_\ell}$, which must be explicitly accounted for
in the expansion. Setting $s_\ell = 2$ for the $m_0 > 0$ case and 
$s_\ell = \ell$ when $m_0 = 0$ leads to perfect agreement with the 
basis-free approach. 

Furthermore, we checked our results for consistency by employing the 
Chapman-Enskog approach presented in Ref.~\cite{Cercignani_book}. Using the properties 
of the RTA collision model, we showed that the Chapman-Enskog method and 
the method of moments are equivalent up to second order. We also showed 
that the discrepancies reported in Ref.~\cite{Jaiswal:2013npa,Panda:2020zhr} 
are due to the omission of second-order contributions in these latter references.

In the context of an ultrarelativistic ideal gas, we computed 
$\gamma^{(\ell)}_{r0}$ and $\Gamma^{(\ell)}_{r0}$ explicitly for 
$\ell = 1$, $2$ and $r = 1$, $2$. We showed that $\gamma^{(\ell)}_{r0}$ 
and all transport coefficients that depend on it, i.e., $\ell_{V\pi}$, 
$\tau_{V\pi}$, $\lambda_{V\pi}$, as well as $\delta_{\pi B}$, $\delta_{VB}$, $\delta_{V\pi E}$, 
diverge with the truncation order $N_\ell$. Even though the coefficients 
$\Gamma^{(\ell)}_{r0}$ also depend explicitly on $N_\ell$, they converge 
towards the basis-free results when $N_\ell \rightarrow \infty$. 

Finally, we validated our results in the context of longitudinal 
waves propagating through an ultrarelativistic ideal gas. Our result $\ell_{V\pi} = 0$ 
for the coefficient responsible for the coupling to the shear-stress tensor 
in the equation for the diffusion current is in perfect agreement with 
numerical simulations of the RTA kinetic equation. 

\begin{acknowledgments}
We thank D.~Wagner, A.~Palermo, P.~Aasha, H.~Niemi, and P.~Huovinen for 
reading the manuscript and for fruitful discussions. 
V.E.A.~gratefully acknowledges the support of the Alexander von Humboldt 
Foundation through a Research Fellowship for postdoctoral researchers. 
The authors acknowledge support by the Deutsche
Forschungsgemeinschaft (DFG, German Research Foundation) through the 
CRC-TR 211 ``Strong-interaction matter
under extreme conditions'' -- project number 315477589 -- TRR 211.
V.E.A.~and E.M.~gratefully acknowledge the support through a grant of the 
Ministry of Research, Innovation and Digitization, CNCS - UEFISCDI,
project number PN-III-P1-1.1-TE-2021-1707, within PNCDI III.
E.M.~was also supported by the program Excellence Initiative--Research
University of the University of Wroc{\l}aw of the Ministry of Education 
and Science. D.H.R.~is supported by the State of
Hesse within the Research Cluster ELEMENTS (Project ID 500/10.006).
V.E.A.~and E.M.~also thank Dr.~Flotte for hospitality and 
useful discussions.
\end{acknowledgments}

\newpage
\appendix

\section{Second-order Chapman-Enskog method}\label{app:CE} 

Recalling the notation introduced in Refs.~\cite{Jaiswal:2013npa,Panda:2020zhr}, 
the distribution function is written as 
$f_\bk = f_{0\bk} + \delta f^{(1)}_\bk + \delta f^{(2)}_\bk$. The correction 
$\delta f^{(i)}$ is obtained as
\begin{equation}
 \delta f^{(i)}_\bk = \left(-\frac{\tau_R}{E_\bk} k^\mu \partial_\mu\right)^i f_{0\bk}\;.
 \label{CE_safari}
\end{equation}
Due to the expansion of the comoving derivative 
$D = \sum_{j = 0}^\infty \varepsilon^j D_j$ in Eq.~(\ref{BTE_CE_rhs}), it is clear that 
$\delta f^{(i)}_\bk$ contains contributions of order $i$, $i+1$, \dots. 
This should be contrasted with the expansion in Eq.~\eqref{CE_df}, 
where $\varepsilon f_\bk^{(1)}$ and $\varepsilon^2 f_\bk^{(2)}$ contain 
solely terms of first and second order with respect to $\varepsilon$, 
respectively, see Eqs.~\eqref{CE_f1} and \eqref{CE_f2}. 

For example, using Eq.~(\ref{CE_safari}) together with Eq.~(\ref{BTE_CE_rhs}) to compute $\delta f^{(1)}_\bk$, 
it becomes clear that it can be written in terms 
of $\varepsilon f_\bk^{(1)}$ and higher-order contributions as
\begin{align}\nonumber
	\delta f^{(1)}_\bk &= \varepsilon f^{(1)}_\bk 
	- \tau_R \sum_{i = 1}^\infty \varepsilon^i D_i f_{0\bk} \\
	&=\varepsilon f^{(1)}_\bk - \varepsilon \tau_R D_1 f_{0\bk} + O(\varepsilon^3) \; ,
\end{align}
where we recall that $\tau_R$ is of the same order as the 
book-keeping parameter $\varepsilon$. The second-order term 
$\delta f^{(2)}_\bk$ can be obtained as
\begin{equation}
 \delta f^{(2)}_{\bk} = \varepsilon^2 f^{(2)}_\bk + 
 \varepsilon \tau_R D_1 f_{0\bk} + O(\varepsilon^3)\;,
\end{equation}
where the second term on the right-hand side makes also a second-order contribution,
being explicitly given by
\begin{equation}
 D_1 f_{0\bk} = 
 f_{0\bk} \bar{f}_{0\bk} \left[D_1 \alpha - E_\bk D_1 \beta 
 - \beta k^{\langle \mu \rangle} D_1 u_\mu\right]\;. 
\end{equation}

The discrepancy between the results derived in the present paper and those 
reported in Refs.~\cite{Jaiswal:2013npa,Panda:2020zhr} arises because  
the second-order contribution $-\varepsilon \tau_R D_1 f_{0\bk}$ to $\delta f^{(1)}_\bk$ 
was neglected in these latter references.
Due to this omission, the resulting distribution function reads
\begin{align}
 \hat{f}_\bk &\equiv f_{0\bk} + (\delta f^{(1)}_\bk +\varepsilon \tau_R D_1 f_{0\bk}) + \delta f^{(2)}_\bk + O(\varepsilon^3)  \nonumber\\
 &= f_\bk + \varepsilon \tau_R D_1 f_{0\bk} + O(\varepsilon^3)\;,
\end{align}
where $f_\bk = f_{0\bk} + \varepsilon f^{(1)}_\bk + \varepsilon^2 f^{(2)}_\bk$.
In the above and henceforth, we use an overhead hat $\hat{f}_\bk$ to denote quantities 
that arise when the $-\varepsilon \tau_R D_1 f_{0\bk}$ term is omitted from $\delta f^{(1)}_\bk$, 
as considered in Refs.~\cite{Jaiswal:2013npa,Panda:2020zhr}.
Using Eqs.~\eqref{CE_exp}--\eqref{CE_rhor} with $f_\bk^{(i)}$ and $\hat{f}_\bk^{(i)}$, we can evaluate 
the difference $\rho_r^{\mu_1 \cdots \mu_\ell} - \hat{\rho}_r^{\mu_1 \cdots \mu_\ell}$ at second order 
as
\begin{equation}
 \rho_r^{\mu_1 \cdots \mu_\ell} - \hat{\rho}_r^{\mu_1 \cdots \mu_\ell} \simeq 
 -\varepsilon \tau_R 
 \int dK\, E_\bk^r k^{\langle \mu_1} \cdots k^{\mu_\ell \rangle} D_1 f_{0\bk}\;.
\end{equation}
In the case of the scalar moments, we find
\begin{align}
 \rho_0 - \hat{\rho}_0 &=
 -\tau_R \frac{G_{20}}{D_{20}}(\Pi \theta - \pi^{\mu\nu} \sigma_{\mu\nu}) \nonumber\\
 & - 
 \tau_R\frac{G_{30}}{D_{20}}(V^\mu \dot{u}_\mu - \nabla_\mu V^\mu) + O(\varepsilon^3)\;,\label{CE_safari_0}\\
 \rho_1 - \hat{\rho}_1 &= -\tau_R (V^\mu \dot{u}_\mu - \nabla_\mu V^\mu) + O(\varepsilon^3)\;,\label{CE_safari_1}\\
 \rho_2 - \hat{\rho}_2 &=
 \tau_R(\Pi \theta - \pi^{\mu\nu} \sigma_{\mu\nu}) + O(\varepsilon^3)\;,
 \label{CE_safari_2}
\end{align}
where \eqref{CE_Dj} was employed to replace $D_1 \alpha$ and 
$D_1 \beta$. 
Since $\rho_1 = \rho_2 = 0$ according to Eqs.~\eqref{delta_n}--\eqref{delta_e},  
it can be seen that $\hat{\rho}_1$ and $\hat{\rho}_2$ will in general not 
vanish. By the same reason, a non-vanishing energy-momentum flow 
$W^\mu = \rho_1^\mu$ appears:
\begin{multline}
 \rho^\mu_1 - \hat{\rho}^\mu_1 = -\tau_R
 \left(\nabla^\mu \Pi - \Delta^\mu_\alpha\nabla_\beta \pi^{\alpha\beta} - \Pi \dot{u}^\mu + \pi^{\mu\nu} \dot{u}_\nu\right) \\
 + O(\varepsilon^3)\;.
 \label{CE_safari_mu_1}
\end{multline}
Equations~\eqref{CE_safari_1}--\eqref{CE_safari_mu_1} show that due to 
second-order inconsistencies, the Landau matching conditions 
(\ref{delta_n}), (\ref{delta_e}) and the Landau frame 
(\ref{Landau_flow}) are no longer satisfied, hence violating the conservation 
of particle number and energy-momentum in the RTA.

The dissipative quantities also show discrepancies,
\begin{align}
 \Pi - \hat{\Pi} &=
 \frac{\tau_R n}{\beta D_{20}} (h J_{10} - J_{20}) (\Pi \theta - \pi^{\mu\nu} \sigma_{\mu\nu}) \\
 & + 
 \frac{\tau_R n}{\beta D_{20}} (h J_{20} - J_{30}) (V^\mu \dot{u}_\mu - \nabla_\mu V^\mu) + O(\varepsilon^3)\;,\nonumber\\
 V^\mu - \hat{V}^\mu &= -\frac{\tau_R}{h}
 \left(\nabla^\mu \Pi - \Delta^\mu_\alpha\nabla_\beta \pi^{\alpha\beta} - \Pi \dot{u}^\mu + \pi^{\mu\nu} \dot{u}_\nu\right) \nonumber \\
 &+ O(\varepsilon^3)\;,
\end{align}
while $ \pi^{\mu\nu} - \hat{\pi}^{\mu\nu} = O(\varepsilon^3)$.
From the above relations, it can be seen that the transport coefficients 
$\delta_{\Pi\Pi}$, $\lambda_{\Pi \pi}$, $\tau_{\Pi V}$, $\ell_{\Pi V}$, 
$\ell_{V \Pi}$, $\ell_{V\pi}$, $\tau_{V\Pi}$, and $\tau_{V\pi}$ are modified 
as follows:
\begin{align} 
 \begin{pmatrix}
  \delta_{\Pi\Pi} \\ \lambda_{\Pi \pi}
 \end{pmatrix} &= 
 \begin{pmatrix}
  \hat{\delta}_{\Pi\Pi} \\
  \hat{\lambda}_{\Pi \pi}
 \end{pmatrix} - \frac{\tau_R n}{\beta D_{20}}(h J_{10} - J_{20})\;,\\
 \begin{pmatrix}
  \tau_{\Pi V} \\ -\ell_{\Pi V}
 \end{pmatrix} &= 
 \begin{pmatrix}
  \hat{\tau}_{\Pi V} \\
  -\hat{\ell}_{\Pi V}
 \end{pmatrix} - \frac{\tau_R n}{\beta D_{20}}(h J_{20} - J_{30})\;,\\
\begin{pmatrix}
  \ell_{V \Pi} \\
  \ell_{V \pi} \\
  \tau_{V \Pi} \\
  \tau_{V \pi} 
 \end{pmatrix} &= 
 \begin{pmatrix}
  \hat{\ell}_{V \Pi} \\
  \hat{\ell}_{V \pi} \\
  \hat{\tau}_{V \Pi} \\
  \hat{\tau}_{V \pi} 
 \end{pmatrix} + \frac{\tau_R}{h}\;.
\end{align} \vspace*{0.5cm}

\bibliography{RTA_coeffs}

\end{document}